\documentclass[manuscript]{acmart} 
\AtBeginDocument{%
  \providecommand\BibTeX{{%
    \normalfont B\kern-0.5em{\scshape i\kern-0.25em b}\kern-0.8em\TeX}}}

\copyrightyear{2023}
\acmYear{2023}
\setcopyright{rightsretained}
\acmConference[IUI '23]{28th International Conference on Intelligent User Interfaces}{March 27--31, 2023}{Sydney, NSW, Australia}
\acmBooktitle{28th International Conference on Intelligent User Interfaces (IUI
'23), March 27--31, 2023, Sydney, NSW, Australia}
\acmDOI{10.1145/3581641.3584094}
\acmISBN{979-8-4007-0106-1/23/03}

\usepackage{tikz}
\usetikzlibrary{arrows.meta}
\usepackage{multirow}
\newcommand{\say}[1]{\textit{``#1''}}

\begin{document}

\title{AutoDOViz: Human-Centered Automation for Decision Optimization}

\author{Daniel Karl I. Weidele}
\email{daniel.karl@ibm.com}
\author{Shazia Afzal}
\author{Abel N. Valente}
\author{Cole Makuch}
\author{Owen Cornec}
\author{Long Vu}
\author{Dharmashankar Subramanian}
\author{Werner Geyer}
\affiliation{%
  \institution{IBM Research}
  \country{USA}
}

\author{Rahul Nair}
\author{Inge Vejsbjerg}
\author{Radu Marinescu}
\author{Paulito Palmes}
\author{Elizabeth M. Daly}
\affiliation{%
  \institution{IBM Research}
  \country{Ireland}
}

\author{Loraine Franke}
\author{Daniel Haehn}
\affiliation{%
  \institution{University of Massachusetts Boston}
  \country{USA}
}

\renewcommand{\shortauthors}{Weidele, et al.}

\begin{abstract}
      We present AutoDOViz, an interactive user interface for automated decision optimization (AutoDO) using reinforcement learning (RL). Decision optimization (DO) has classically being practiced by dedicated DO researchers \cite{nelder1965simplex} where experts need to spend long periods of time fine tuning a solution through trial-and-error. AutoML pipeline search has sought to make it easier for a data scientist to find the best machine learning pipeline by leveraging automation to search and tune the solution. More recently, these advances have been applied to the domain of AutoDO \cite{marinescu2022automated}, with a similar goal to find the best reinforcement learning pipeline through algorithm selection and parameter tuning. However, Decision Optimization requires significantly more complex problem specification when compared to an ML problem. AutoDOViz seeks to lower the barrier of entry for data scientists in problem specification for reinforcement learning problems, leverage the benefits of AutoDO algorithms for RL pipeline search and finally, create visualizations and policy insights in order to facilitate the typical interactive nature when communicating problem formulation and solution proposals between DO experts and domain experts. In this paper, we report our findings from semi-structured expert interviews with DO practitioners as well as business consultants, leading to design requirements for human-centered automation for DO with RL. We evaluate a system implementation with data scientists and find that they are significantly more open to engage in DO after using our proposed solution. AutoDOViz further increases trust in RL agent models and makes the automated training and evaluation process more comprehensible. As shown for other automation in ML tasks \cite{weidele2020autoaiviz, lee2020human}, we also conclude automation of RL for DO can benefit from user and vice-versa when the interface promotes human-in-the-loop.
\end{abstract}



\begin{teaserfigure}
  \includegraphics[width=\textwidth]{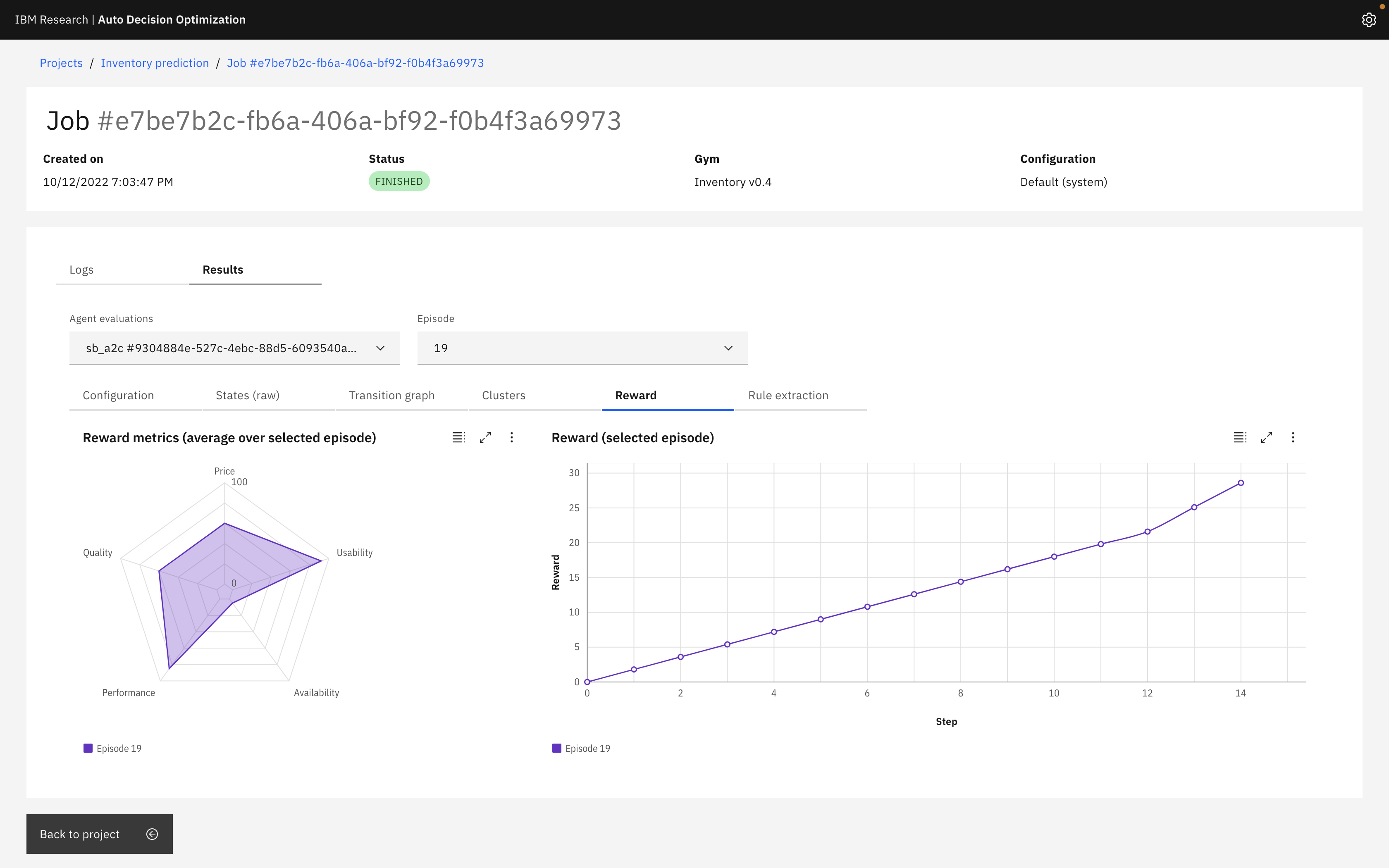}
  \caption{Reinforcement learning agent evaluation in AutoDOViz.}
  \label{fig:teaser}
\end{teaserfigure}

\received{15 October 2022}
\received[revised]{03 February 2023}
\received[accepted]{10 February 2023}

\maketitle
\section{Introduction}

Decision optimization (DO) is a discipline tightly integrated with business domain users to enable better, faster, and more efficient decision-making. DO systems aim to recommend single decisions or present multiple alternatives to users and improve continuously while in production. DO finds application in virtually every industry and business practice that involves decision-making and optimization of resource allocation, such as uncertain demand in inventory networks or intelligent scheduling of heating, ventilation, and air conditioning in building control \cite{wei2017deep}. However, DO is a reasonably specialized, knowledge-oriented field that requires expertise in operations research and mathematical optimization and a deep understanding of specific business processes. This combination is challenging to find in individuals, so tight integration across teams or specifically tailored software solutions is necessary to succeed in DO projects. Recent advances in reinforcement learning (RL) have created new opportunities to add to the DO toolbox \cite{marinescu2022automated}. RL gyms flexibly encode problem settings in which RL agents decide actions to optimize a given reward function. Furthermore, advances in automation have shown how the process of designing high-quality machine learning pipelines accelerates the democratization of Artificial Intelligence (AI) from machine learning researchers and engineers towards data scientists and business users~\cite{jin2019auto, ledell2020h2o, olson2016evaluation, feurer2015efficient, das2020amazon, mukunthu2019practical}. 
In this paper, we want to take the next step towards a \emph{user experience}\footnote{Throughout this paper we use the term \emph{user experience} in a broader sense for the system design, its implementation and usability but would like to point the reader in acknowledgement to more granular differentiation as suggested in previous work \cite{hassenzahl2006user,desmet2007framework,law2007towards}.} for automation of decision optimization with RL to accelerate democratization from DO researchers to data scientists and business users. Our goal is to empower a broader set of people to develop optimization models and make automated decision-making more accessible overall. A fundamental task on this path is to precisely understand the decision optimization process, the tools used, and associated challenges to identify opportunities and scope for automation and support. Additionally, characterizing the range of business domain stakeholders and users, their skills, goals, and expectations can help design a more practical and effective solution. We, therefore, conducted extensive semi-structured one-on-one interviews with DO practitioners and business domain experts to provide insights into these questions. 
Regarding business experts, we aimed to characterize potential stakeholders, such as their ability to use such a system as well as their needs, goals, and expectations, which can help design a solution that is more practical and effective as well as enhancing the democratization of AI for our target group, such as the general data scientist. Based on our requirements, we propose a novel system design for automated decision optimization using reinforcement learning, AutoDOViz. AutoDOViz builds on the AutoDO core engine~\cite{marinescu2022automated}, which leverages search techniques to automate algorithm selection and hyperparameter optimization. In AutoDOViz, we bring together novel UI interfaces to simplify the problem formulation process, leveraging the AutoDO capabilities to generate optimized agents and, finally, policy visualizations and metrics to enable better communication between the agent designer and the domain expert to align the end solution with domain expert needs and inputs. 
We rigorously test our design in an evaluation study with data scientists and report our findings, which also leads to future work.



\paragraph{The contribution of this paper is 4-fold.} After reviewing related work in section 2, we present insights from expert interviews with (1) decision optimization researchers as well as (2) business domain consultants, which lead to system design requirements in section 3. Section 4 provides details about our (3) implementation of a user experience for an automated DO system with RL (\textit{"AutoDOViz"}). Lastly, section 5 contains the evaluation of AutoDOViz in the form of (4) a user study with data scientists. We discuss our results in section 6 before drawing conclusions and suggesting directions for future work in section 7. 

\section{Related Work}

\subsection{Decision Optimization and Reinforcement Learning}
In decision optimization, the goal is to find the best solution for an optimization problem according to a given criterion among a set of possible solutions. Therefore, the use of optimization algorithms as general step-by-step procedures plays a crucial role. Optimization algorithms are exact if they can find the optimal solution to the problem. They are heuristic if they propose an appropriate solution that is not necessarily optimal in the solution set~\cite{barbati2012applications}. Optimization problems are commonly found in scheduling problems, transportation, logistics, or supply chain planning \cite{barbati2012applications}. In general, decision optimization can benefit from creating interactive methods, especially dynamic interaction between the user and the system~\cite{chiu2008hyper}. Interaction with end-users makes the optimization procedure more effective by enriching the selected optimization model or informing the decision maker regarding available solutions proposed by the optimization system~\cite{meignan2015review}. Using machine learning for combinatorial optimization to learn an approximate relation between decisions and their impact on the system plays a crucial role in current research. Such systems suggest optimal decision outcomes in complex real-world settings by taking advantage of the recent developments in artificial intelligence~\cite{lombardi2017empirical}. Recent literature confirms that agent-based approaches can successfully cope with a large spectrum of optimization problems. Agent-based computing has computational advantages as problems can be divided into multiple sub-problems, which can then be solved by massive parallelization (e.g., involving GPU). This is preferable for large problem spaces \cite{barbati2012applications}. Earlier approaches also already proposed solving optimization problems, such as dynamic job shop scheduling, by using reinforcement learning agents~\cite{aydin2000dynamic}.

Reinforcement learning \textit{(RL)} or deep reinforcement learning \textit{(DRL)} have become prominent techniques in machine learning to address sequential decision-making problems by training autonomous agents \cite{sutton2018reinforcement, kaelbling1996reinforcement}. 
A trained model is usually referred to as an independent agent that interacts in a particular environment with specific predefined actions. RL uses a trial-and-error approach to learn an optimal policy or behavioral strategy~\cite{barto2004intrinsically}. The RL algorithm surveys the optimal action space given observed feedback on decisions, for example, a reward from the environment, to learn which actions are beneficial in a given state. Primarily, RL's success originates from RL agents' ability to solve decision-making problems in complex and uncertain environments. RL is popular in various fields such as autonomous driving, robotics, health, finance, smart grids, video games, and education \cite{li2017deep}. Other applications of RL include inventory management with multiple stages and suppliers with strict lead times under demand uncertainty or control problems such as autonomous manufacturing operations with resource allocation constraints. RL can also reduce the need for repetitive human interaction by decreasing manual work like calibrating or monitoring a system, allowing systems to adapt faster to changes~\cite{saldanha2019relvis}. RL agents aim to maximize the total expected reward by learning an optimal policy. 

However, RL agents might lack semantic knowledge on the current goal, i.e., \textit{why} some actions and elements are preferable over others~\cite{sequeira2019interestingness}. The agent should be able to explain their reasoning and decision-making process to users, which could increase the user's trust towards the RL agent. Explainable Reinforcement Learning (XRL) is a form of Explainable Artificial Intelligence (XAI), focusing on the importance of providing explanations to gain the trust of users~\cite{dzindolet2003role, chatzimparmpas2020state, puiutta2020explainable, mohseni2021multidisciplinary}.
Several methods have been proposed to provide greater transparency for users
~\cite{du2019techniques, vellido2012making, doshi2017towards}. With the increasing use of intelligent agents, it is essential to introduce methods to uncover agent behaviors and limitations. For example, algorithms provide users a behavioral summary of the agent in different situations by extracting critical trajectories from simulations of the agent~\cite{amir2018highlights}. \citet{gajcin2022contrastive} further suggests that understanding the differences in strategies between RL policies is required to enable users to choose between those policies. 

\subsection{Visual Analytics for Reinforcement Learning}
Next to XRL, visual analytics is a promising technique for making the RL training and evaluation process more understandable and trustworthy. 
The use of visual analytics tools and interactive user interfaces provides insights on decisions and representations~\cite{hohman, yuan2021survey}. Within RL, 
ReLVis~\cite{saldanha2019relvis} is a visual analytics tool designed to help data scientists keep track of RL experimentation. It visualizes relations between hyperparameter changes and outcomes, such as behaviors associated with specific rewards and how agent behavior evolves. DRLViz~\cite{jaunet2020drlviz} is an interface to interpret internal memory of an agent. DRLIVE~\cite{wang2021visual} is a visual analytics system for RL that tracks agents through interactive visualizations and further diagnoses the DRL model by perturbing its inputs. PolicyExplainer~\cite{mishra2022not} is a visual analytics interface that allows users to query an autonomous agent directly. The tool extracts data from a trained agent, summarizes the optimal policy, analyzes states and expected rewards, and visualizes a trajectory to see the agent's progression. Moreover, tools for the visualization of specific RL algorithms have been presented. For example, DQNViz~\cite{wang2018dqnviz} visualizes states of Deep Q-networks (DQNs) during the training of a deep RL agent.
DynamicsExplorer~\cite{he2020dynamicsexplorer} diagnoses an RL policy under different dynamic settings. Direct visualization of the internals of RL agents has been discussed in past research for games~\cite{greydanus2018visualizing}, including dimensionality reduction, reward curves, or Jacobian saliency maps~\cite{zahavy2016graying, mnih2015human}. However, they all point out the growing need for human-interpretable explanations of deep RL agents and difficulty for non-experts to interpret. Especially in real-world scenarios, where design problems consist of more than one conflicting or cooperative objective (\textit{"multi-objective optimization problems"}), visualization can be challenging~\cite{chiu2008hyper}. Overall, more research and exploration on how to connect RL with visualization is required~\cite{wang2021survey}.


\subsection{User Experiences for Automation in Machine Learning}

Novel methods have been developed to automate parts of the data science process, from data pre-processing, feature engineering, and model selection to hyperparameter optimization \textit{(HPO)}~\cite{zoller2021benchmark, kanter2015deep, khurana2016cognito, lam2017one, wang2019atmseer}. While much work has been shown to help solve specific problems, only a few can run in a fully automated and interactive way to enhance usability and decrease manual work. Mainly HPO relies on data scientists' experience and the dataset itself. Therefore, automated methods for HPO are also an active area of research in AutoML~\cite{gambella2021optimization}. Recently, new research has emerged to automate the full artificial intelligence lifecycle, so-called Automated Artificial Intelligence (\textit{AutoAI}) or Automated machine learning \textit{(AutoML)}. The goal is to enable domain experts to automatically build ML applications without extensive statistics or AI knowledge. Furthermore, such systems should reduce repetitive, manual workloads for data scientists and shift their time spent on summarizing and controlling visualization techniques \cite{weidele2020autoaiviz, weidele2019conditional}. Commercial AutoML and AutoAI products have been released as fully managed machine learning services. Examples are Amazon SageMaker\cite{das2020amazon}, Azure AutoML\cite{mukunthu2019practical}, H2O AutoML~\cite{ledell2020h2o}, Google AutoML \cite{bisong2019google} and IBM AutoAI~\cite{wang2019human, weidele2020autoaiviz, shah2021autoai}. Additionally, open-source software packages are available, such as AutoKeras~\cite{jin2019auto}, TPOT~\cite{olson2016evaluation}, Auto-Sklearn~\cite{feurer2015efficient} and LALE~\cite{hirzel2022gradual}.
Recently, \citet{marinescu2022automated} have proposed an automation engine towards data- or knowledge-driven dynamic optimization problems using reinforcement learning.


\section{Explorative Interviews}

\subsection{Decision Optimization Researchers' perspective}

Semi-structured interviews were conducted to collect individual opinions and experiences of DO  practitioners while allowing flexibility for deeper discussions or exploration of people’s views or narratives. Initial research questions were framed as prompts to guide the interview process but were not strictly adhered to formally or linearly. Instead, the interviews evolved into free-flowing conversations where participants delved into their subjective experiences and reflected on topics of interest. 

\subsubsection{Participants}

These interviews were conducted with six people with extensive applied industry experience in DO.
Five of the six people we interviewed had over 20 years of experience in optimization and vast experience developing and consulting optimization solutions for clients. Only one was a data scientist but had practical experience working on optimization problems. All interviews were conducted remotely and recorded using standard commercial video conference software with an automatic transcribing facility. The interviews took about 60 minutes on average, giving rich insights into our research questions. 
We know that this sample size is limited in terms of the number of participants and range of optimization experience. However, since our interview participants are experts with extensive experience in optimization, their views can be considered reliable and valid for the early insights we were looking for. Note that optimization is an extremely niche domain. One of the aims of AutoDO is to increase the accessibility of DO solutions to a broader user base that may not have significant OR/DO expertise. So, it is meaningful to engage with DO experts to design for the non-DO experts we aim for as target users.

\subsubsection{Analysis}
Thematic analysis is a popular method to uncover meanings from qualitative descriptions of experiences and contexts. Using thematic analysis, the interview data was analyzed to find patterns and themes of interest. The initial set of research questions guided the coding of this data and helped to formulate some actionable insights. 

\subsubsection{Results}
The interviews' findings and implications for an automated solution are structured by the various themes and dimensions we identified during the analysis. 

\paragraph{Domains}
Our interview participants reported first-hand experience across various domains like energy, transportation, aviation, retail, finance, education, defense, semiconductors, telecommunication, and healthcare. This emphasizes the relevance of decision optimization and its potential to help businesses drive timely and efficient decisions. Optimization models are required for tasks like determining production capacity, planning inventory, supply chain routing, operations, and maintenance scheduling, workforce planning, and manufacturing and warehouse management. The challenge is that these models are sensitive to a given business context, and they must be customized considering the constraints and objectives of each business or client. This makes it difficult to think of a generic DO model that is portable across contexts and domains. However, it appears that various optimization tasks can be broadly categorized into the following four categories based on factors like time and precedence: resource assignment (1), selection \& allocation (2), supply \& demand planning (3) and scheduling (4).

For automated DO, this simplifies the problem space into a categorization that can enable building resources aligned to these four generic templates \textbf{[R1]}. For example, visualizations that are an important mode of communicating results with the stakeholders can be built to incorporate the essential characteristics and requirements for each category instead of customizing each business problem \textbf{[R2]}.

\paragraph{User Identification and Characterization.}

Various internal and external stakeholders can be identified based on their needs and goals in generating a DO model. The interview data's roles, titles, and personas ranged from business representatives, business analysts, data engineers, data scientists, planners, quantitative analysts, developers, IT specialists, business consultants, and salespeople to one or more optimization experts. This is a broad spectrum of technical and non-technical expertise with clearly different but highly dependent roles in the workflow. 
This implies that any automated solution should ideally be collaborative if it intends to be a one-stop workbench for creating an end-to-end DO model \textbf{[R3]}. This, in turn, necessitates the characterization of skills and identification of goals explicitly when designing the user interface \textbf{[R4]}. One can also consider creating levels of complexity in the solution interface so that supported personas can easily navigate between more transparent and detailed or complex views of the pipeline. A hierarchical granularity or visibility into the execution flow may well serve the requirements of the diverse stakeholders involved at various stages of the solution process. Given the complexity of designing and implementing an end-end solution, it is likely that initial versions of automated DO can only support model development with explanations and a provision to import and export process modules (like input and output) for communicating with the broader set of stakeholders.

\paragraph{Tools.}
We found that DO practitioners use a range of tools, often building customized UIs for clients. The popular modelling languages used are OPL, Python, Java and even C++ with a clear preference towards OPL by our expert sample. Specifically, OPL seemed to be the choice for building quick prototypes for client demos. 
From informal notes, whiteboard discussions, excel sheets, to the use of special and general purpose modelling languages and enterprise software suites – there seems to be a lack of a powerful one-stop workspace that offers end-to-end workflow support. A need for interactive visualizations at various levels of granularity was a key missing piece of technology identified in our data. 
It is worthwhile to explore designing a solution that supports an end-end workflow providing users with an integrated framework for ease and usability \textbf{[R5]}.

\paragraph{Data Scientists as DO Practitioners.}

Data scientists are being considered as the primary target users for AutoDO. Even though most data scientists have a mathematical background they generally use statistical or machine learning techniques to derive insights from data or build ML models. Since AutoDO is also data-driven, the data scientists are a natural choice to be potential end-users. 
Our interviews with DO experts sought to understand potentials for introducing DO as a method to data scientists and the challenges they might face. While it is uncommon for data scientists to solve DO problems, we did have one participant (ID 1) who had some experience in working on optimization problems. Interestingly, the challenges he faced had less to do with the modelling part but were mostly related to approaching the DO problems, framing or scoping them and explaining the solution to the clients. This echoes exactly with views of participant ID 2 who believes that optimization is just another technique that a data scientist can use but will require an understanding of the business and what one needs to model to be able to formulate the right problem. Understanding the business problem and not blindly taking data to try different things is what differentiates a data scientist from an optimization expert for ID 4. As ID 3 puts it – “DO is a different way of thinking that requires mapping between business to decision variables and constraints“ and a lot depends on what would motivate a data scientist to consider DO for problem-solving (ID 4). This is an interesting aspect also mentioned by ID 5 saying that a “data scientist would be completely lost if they don’t already know that the problem is an optimization one” or even ID 6 mentioning that “Data scientists do not uncover optimization opportunities”. Highlighting the importance of business knowledge, ID 6 quantifies the role of a data scientist as relying about 70\% on data and 30\% on the business process and vice-versa for an optimization expert. For the former the question is about how to build an ML model from given data and infuse it into the existing process; while for an optimization expert a deep understanding of the business process and scoping it determines the data requirements. It is the thinking and refinement of the specified business problem within the given constraints that distinguish the optimization expert from the typical data scientist. 
To consider data scientists as target users for AutoDO depends a great deal on which aspects of the DO workflow are incorporated into it. Scoping or problem specification and model evaluation are the most challenging phases and in the real-world are conducted in close collaboration with other stakeholders like business analysts or subject matter experts. 
In the standard DO process, testing and validation of the solution on KPIs or against various scenarios and use-cases determines its acceptability. A data scientist in the role of an optimization solution builder has neither the mandate nor the proficiency to take this decision individually.

\paragraph{Leveraging Automation and AI}

We asked our participants about how automation or AI could help in overcoming some of the challenges and bottlenecks they currently face. We find that developers need tools to monitor performance and understand execution flow, while consumers need to trust and have confidence in the solution \textbf{[R6]}. Therefore, explanations and interactive visualizations that help monitor the execution and enable communication between all the stakeholders was unanimously called out. 
Testing the solution on historical data or using multiple scenarios, replaying decisions, ML guided tuning of solution, using visualizations to uncover bottlenecks, graphical interactive monitoring of solution, and gathering of insights are some areas where our participants desired tooling or automation support. Interestingly, there is a strong requirement for interactive UIs and rich visualizations in the community and AutoDO has an opportunity to make a convincing contribution in this space.


\subsection{Business Domain Consultants' perspective}
We further aim to improve AutoDO’s relevance to real-world use cases by conducting semi-structured interviews with business consultants. We sought consultants with deep knowledge in particular industries, rather than generalists, to learn about potential optimization problems within that specific industry. 

The interviews were based on the following six semi-structured questions:
\begin{enumerate}
    \item What is your primary industry?
    \item What kinds of clients do you work with in this industry?
    \item What are main problems these clients face?
    \item Have you seen integration of AI in this industry? What does it do?
    \item Which problems in this industry cannot be solved with AI?
    \item Reinforcement Learning trains agents via reward and punishment for actions and can be good at making discrete decisions such as driving instructions for self-driving cars, moves in games such as chess and checkers, or recommendations in the stock market. Can you think of decisions in your industry that would be particularly helpful to automate?
\end{enumerate}

The interviews were scheduled for 30 minutes in length, and conducted remotely over video conference during the consultant's workday with two interviewers in attendance.

A total of eight interviews were conducted with consultants (henceforward, 'specialists') specializing in Advertising, Agriculture, Automotive, Government (two consultants), Industrial Manufacturing, Oil, and Retail. Everyone except of the Oil specialist was hand-selected from a staff list of 547 global consultants. 
Our interviews consisted of three women and five men, ranging in age from early-career to mid-career. Information on age or race was not solicited. 


\subsubsection{Summary of findings} 
The \textbf{Advertising} specialist was an early-career generalist consultant who 
communicated that many of the core tasks in advertising are still time-consuming manual processes, and approximated the breakdown of big data, AI, and human input for each stage of the digital advertising work cycle. They believed that algorithms are efficient at optimizing many decisions, but that humans were necessary to drive high-level strategy and focus algorithms on particular tasks. They gave specific examples of automated processes that they believe required human input to function optimally. Based on our explanation of RL in question six, they believed that RL could be most useful for identifying the types of content that resonate with people.


The \textbf{Agriculture} specialist was a mid-career consultant 
who grew up on a farm, had a formal education in Agronomy (agricultural consulting), and spent their entire pre-consulting career with agriculture firms. 
They claimed that farmers were generally distrustful of new technology,
but that application of Big Data was increasingly common in agriculture. For example, farmers are now able to purchase crop insurance based on conditions such as the maximum outside temperature during a particular period, which is calculated using advanced underwriting models. 

The \textbf{Automotive} specialist was a mid-career consultant 
with additional background in retail consulting.
They identified current trends in the automotive industry and felt RL's impact in the automotive industry were mainly related to self-driving cars as referenced in question six. When challenged to propose applications unrelated to the driving experience, they suggested RL could help manufacturers identify how to configure future cars according to consumer preferences.


The first \textbf{Government} specialist 
had spent over a decade consulting for multiple government agencies. They have observed the demand for data within government agencies is changing. Particularly, they described how legacy systems had been adapted to meet modern demands for information access and transparency. They identified handling requests for security clearances as a potential application for RL. 
The second Government specialist had worked for various federal agencies, including defense and security agencies, before their time as a consultant. They believe that distributing and optimizing the use of resources across locations and agencies is the primary problem for the government organizations they had interacted with.
They presented the potential issue of getting security clearance for AI models and programs. Their proposed RL use cases were for generating government reports and testing new applications.

The \textbf{Manufacturing} specialist had prior experience at a coal power and mining firm, and had primarily consulted for coal, oil, and large pharmaceutical companies during their consulting career. The key insight from this conversation was the importance of constant analysis of the entire supply chain. 
They believed that much of the technical innovation in the manufacturing industry is related to accurate modelling and analysis of when and how machines might fail, and that one of the biggest current problems in manufacturing is insufficient information. This specialist believed that AI could have the biggest impact on manufacturing by improving efficiency, and proposed use of RL to monitor the consumption, use, and disposal of raw materials according to material-specific considerations.

The \textbf{Oil} specialist was a geologist by training, and had spent their consulting career working with oil companies in and off the coast of Texas. 
They identified pieces of the drilling process that are currently trial-and-error rather than data-driven, and offered that proper modelling and responses to models could significantly increase the longevity and output of wells. They also shared that oil infrastructure was an interconnected superstructure of equipment from many manufacturers, and navigating the supply chain required constant real-time communication between many different stakeholders.

The \textbf{Retail} specialist had experience with
some of the world’s largest big-box retailers and responsible for overseeing the creation and acquisition of multiple e-commerce retail organizations. They shared thoughts on the emerging issues presented by the demand for retail delivery, the current state of warehousing and changing philosophies on stocking, and how to render the in-store experience online.

\subsubsection{Discussion}

The consultants 
offered useful and nuanced insight into how AI is currently employed in their respective industries and potential future applications. We found that questions related specifically to RL yielded less actionable results than general discussion about AI, which leads us to believe that business people have a good sense for AI's capabilities, but a limited sense of the differences between training methods. It seems useful for a tool to provide insights on the RL process to end users, so that it becomes clear how the training process yields the trained model \textbf{[R7]}.

The discussion of specialized problems within particular industries highlighted the usefulness of breaking down templates of DO problems by industry. An interface that helps users identify the type of business their problem is centered on would help them more easily locate relevant solutions \textbf{[R8]}. We also learned of types of non-industry-specific business processes that are very relevant to multiple or all industries. 
These kinds of general business DO problem templates should also be made available within industries, even when they are shared among different industries \textbf{[R9]}.


\subsection{Design Requirements}

Capturing experts' needs and interests from our detailed semi-structured interviews, table \ref{tab:requirements} summarizes requirements for human-centered automation for DO with RL.

\begin{table}[H]
    \begin{center}
    \caption{System requirements for human-centered automation for decision optimization.}
    \label{tab:requirements}
    \begin{tabular}{l l}
        \hline
        \textbf{R1} & Generic templates which align with common categories of DO problems. \\
        \textbf{R2} & Visualizations for DO problem categories rather than for each business problem. \\
        \textbf{R3} & Encourage collaboration between stakeholders. \\
        \textbf{R4} & Characterization of skills and identification of goals for supported personas. \\
        \textbf{R5} & Support end-to-end workflow with an integrated framework for ease and usability. \\
        \textbf{R6} & Promote trust and confidence in the solution. \\
        \textbf{R7} & Explain the RL training process for non-technical users. \\
        \textbf{R8} & Classify templates by industry for ease of location. \\
        \textbf{R9} & Provide templates for general business problems shared by multiple industries. \\
        \hline
    \end{tabular}
    \end{center}
\end{table}



\section{System Design}

\subsection{Integrating Core Engine and User Interface Elements}

AutoDO \cite{marinescu2022automated} is an end-to-end system for automated reinforcement learning using a bi-level nested search to optimize RL pipelines. More specifically, in the first level, the system admits pipelines with different flavours of RL, namely online and offline methods combined with model-free and model-based approaches. In the inner level, it further admits a mixed discrete and continuous search space to tune the RL agent hyperparameters as well as the neural architecture choices. The system employs different search strategies to traverse this complex search space such as random search, Bayesian optimization or a more recent multi-fidelity strategy based on limited discrepancy search \cite{kishimoto2022aaai}. 

The input to the system consists of a collection of RL agents and either a gym-compatible environment or a dataset containing tuples $(s, a, r, s')$, where $s$ and $s'$ represent the current and respectively the next state of the underlying Markov decision process, $a$ is the current action and $r$ is the reward. Optimization problems can be provided in the form of RL environments based on the open-source toolkit OpenAI Gym~\cite{brockman2016openai}, a Python programming interface. The RL agent then acts on the OpenAI Gym in a series of episodes, where each episode starts with a user-defined initial state followed by interactions with the environment until a user-defined termination criteria is reached. The performance of RL agents is assessed via user-defined a reward function. The output consists of a the set of top $k$ pipelines produced by the joint pipeline and hyperparameter optimizer and representing the best performing agents with their hyperparameters.

\begin{figure}
    \centering
    \includegraphics[width=1\linewidth]{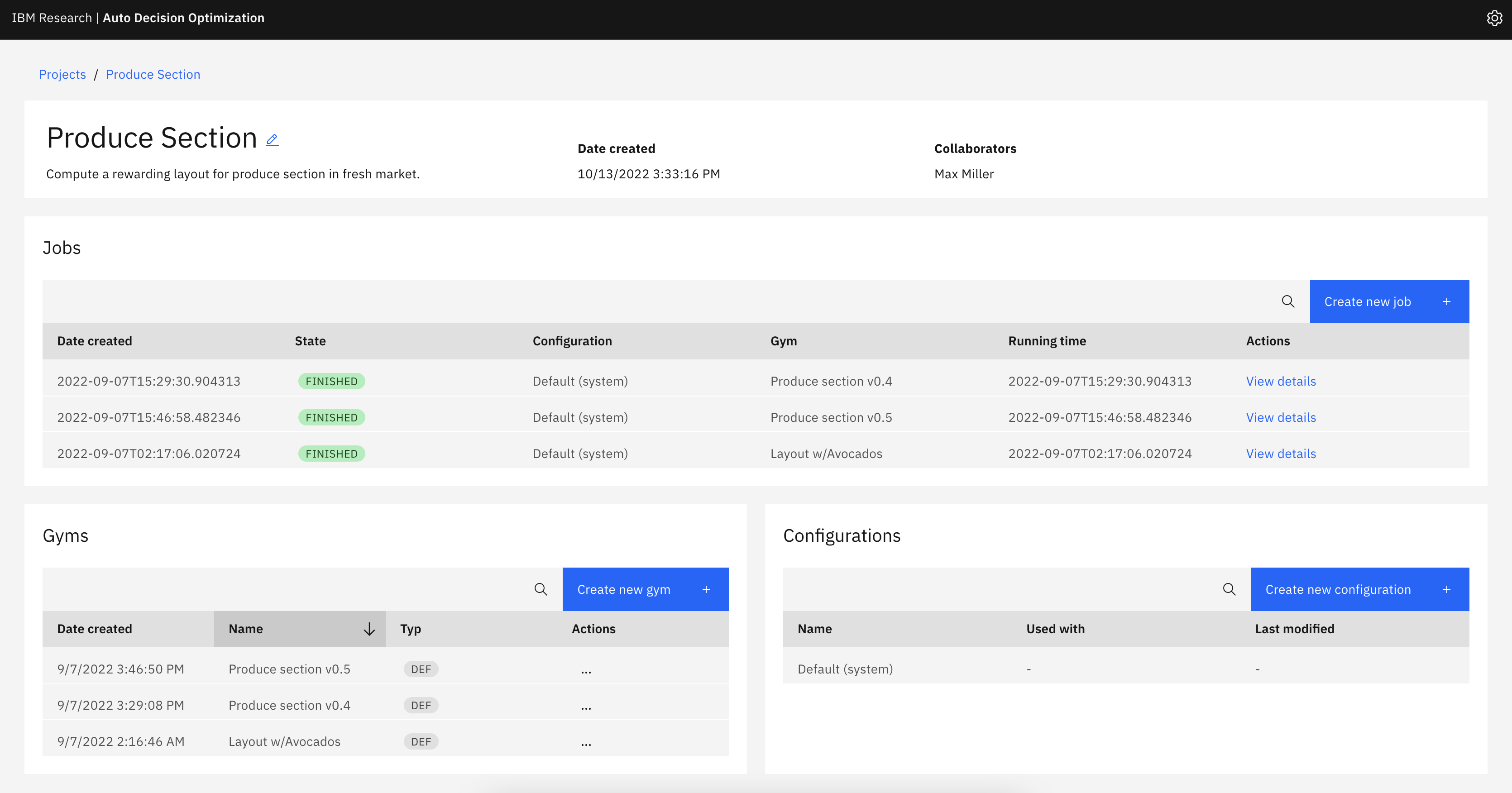}
    \caption{\textbf{Project dashboard in AutoDOViz} Users can manage gyms, engine configurations and trigger RL agent training.}
    \label{fig:proj_dashboard}
\end{figure}

Based on this specification of the core engine, we identify three entities that need to be managed by the user of AutoDOViz: gyms (or environments), engine configurations, and resulting jobs (or executions). Regarding requirement \textbf{R3}, we also propose organizing these entities in \textit{projects}, which can then be shared among stakeholders. For example, a business domain user does not necessarily need to be able to modify the gym implementation. However, they might want to trigger a gym execution 
and assess high-level visualizations. While gyms are only accessible within a project, configurations of the AutoDO engine can be shared across projects. Thus, users can bring along configurations that work well in other cases when they move on to a new project. Figure \ref{fig:proj_dashboard} shows the project dashboard with a clear layout and a simple table view for each of the three entities.

\begin{figure}
    \centering
    \includegraphics[width=1\linewidth]{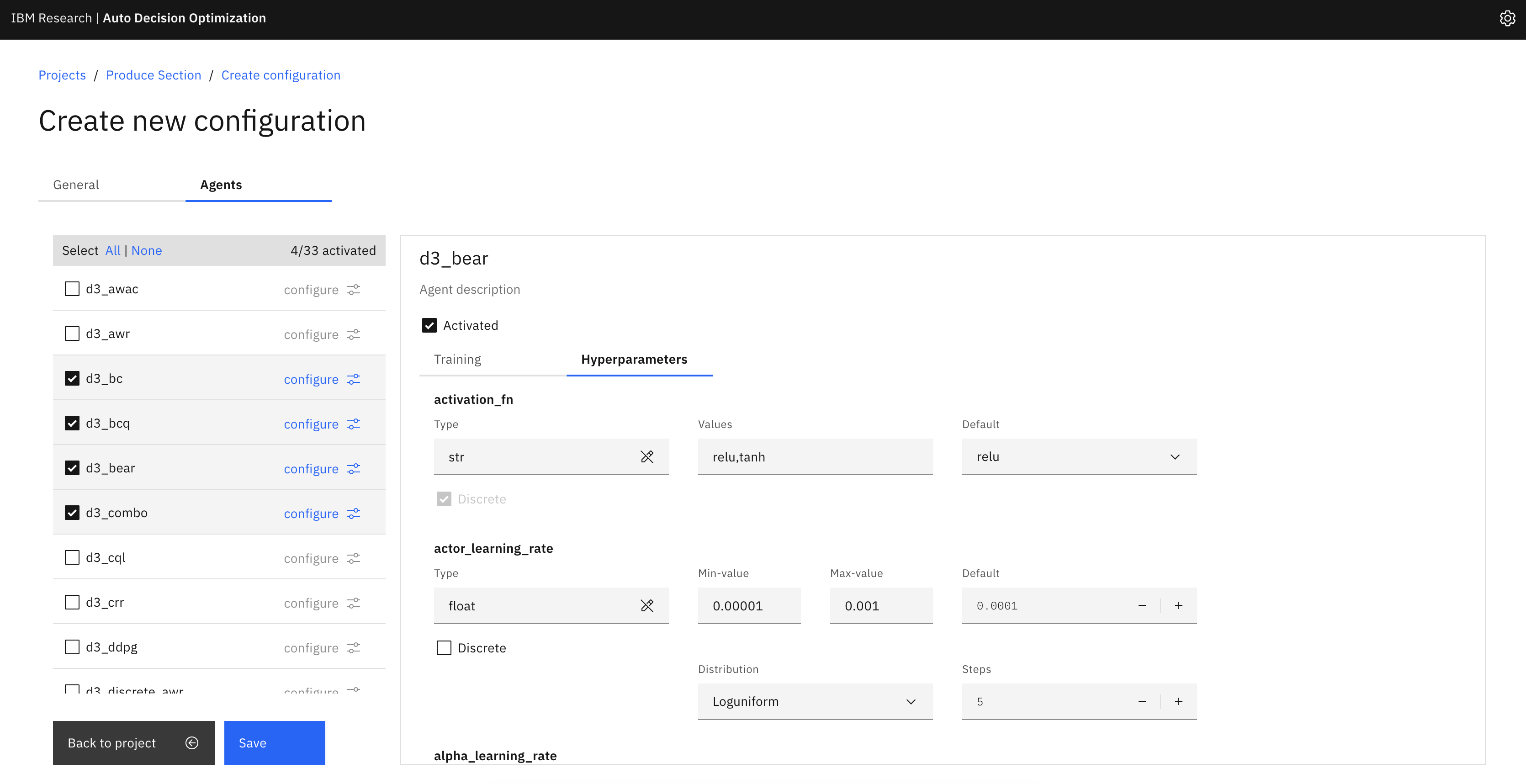}
    \caption{\textbf{Optional AutoDO engine configuration in AutoDOViz}. Users can toggle individual agents and set hyperparameter search constraints.}
    \label{fig:configuration}
\end{figure}

Figure \ref{fig:configuration} depicts the configuration wizard of AutoDOViz. The left hand side pane shows the list of selected RL agents. The right hand side shows the tunable hyperparameters of a particular agent (e.g., \texttt{d3\_bear}). For each hyperparameter we specify the type, possible values if discrete or the range if continuous, as we well as the default value. The result of this configuration phase is the definition of the joint pipeline and hyperparameter search space traversed by the AutoDO engine. 

We handle the user interface elements for gyms in more detail throughout sections \ref{section:composer} and \ref{section:catalog}, as well as for job results/executions in sections \ref{section:stateviz} and \ref{section:policy}.

\subsection{(Remote) Execution with Streaming Architecture}

\begin{figure}
    \centering
    \includegraphics[width=0.7\linewidth]{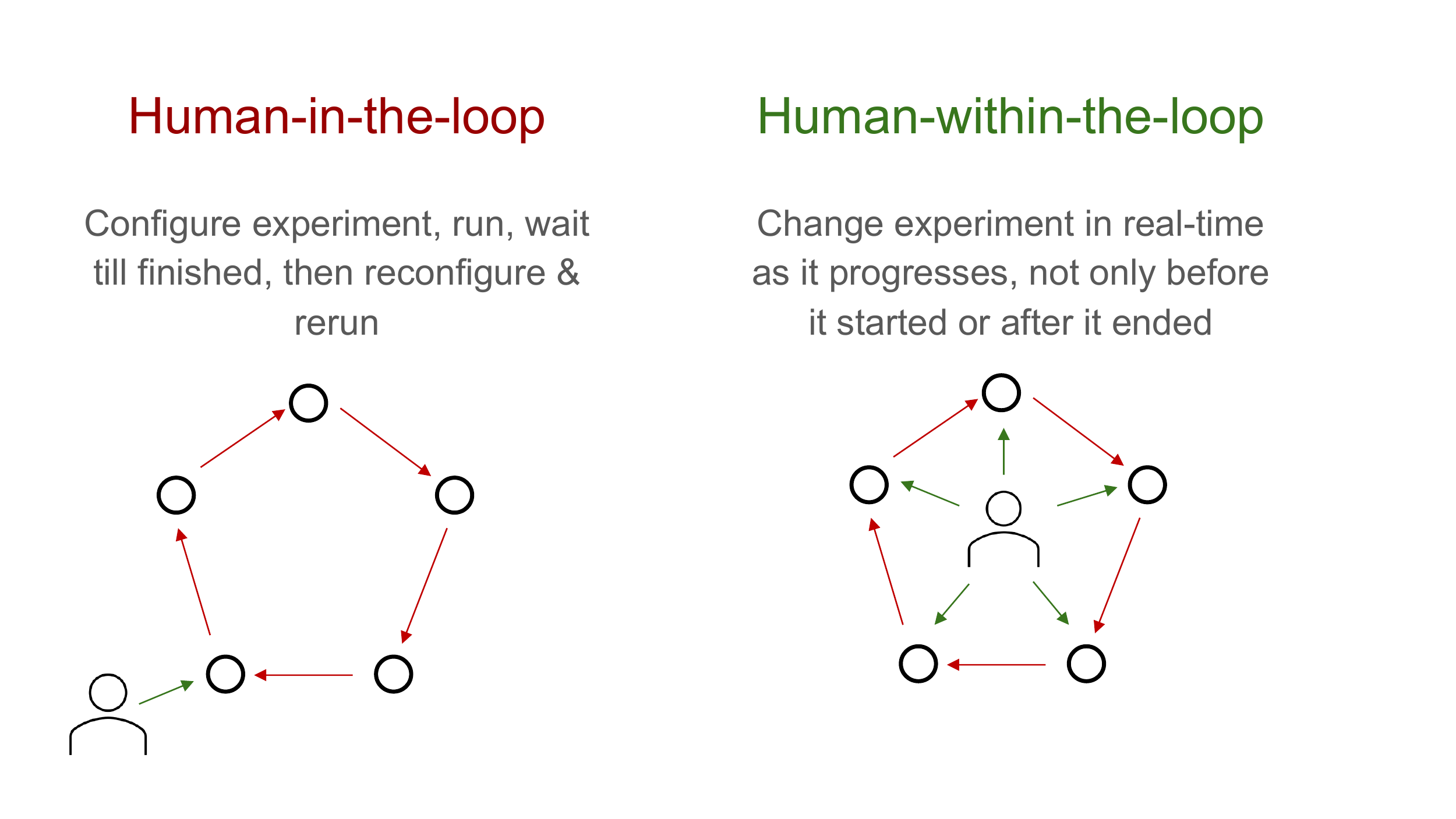}
    \caption{\textbf{Human-within-the-Loop.} Instead of round-based interaction at the beginning and end of each cycle (\textit{left}), we aim to include the user in automation in real-time to promote high interactivity and usability.}
    \label{fig:humanloop}
\end{figure}

User experiences for automation often incorporate an explicit Human-in-the-loop (HitL) mechanism \cite{wang2019human, weidele2020autoaiviz, gil2019towards, lee2020human, meignan2015review}. Users of HitL optimization systems are part of the optimization process after every turn of the automation (see Figure~\ref{fig:humanloop}). In our implementation, specifically under consideration of \textbf{R5}, we aim to evolve this approach towards a more involved Human-\textit{within}-the-loop (HwtL) process (Figure~\ref{fig:humanloop} right). With HitL being an important aspect of designing automated optimization systems, designing more towards HwtL should lead to beneficial progress of an intelligent user interface ~ \cite{meignan2015review}. Since HwtL constitutes a very high technical challenge, in this prototype we currently only support real-time \textit{tracking} of the AutoDO engine, which means the user already has full transparency of the process at any step of the process. As we will now describe details of our streaming architecture, we would like to highlight that full real-time \textit{manipulation} of the engine is \emph{not yet} part of this work, but will benefit from the same architectural design in a future iteration of the software. For example, users will then also be able to add/remove RL agents or modify hyperparameter search constraints in real-time during the execution of the engine to focus computational resources or to allow for quick intermediate \emph{What-If?} scenarios.

\begin{figure}
    \centering
    \includegraphics[width=0.5\linewidth]{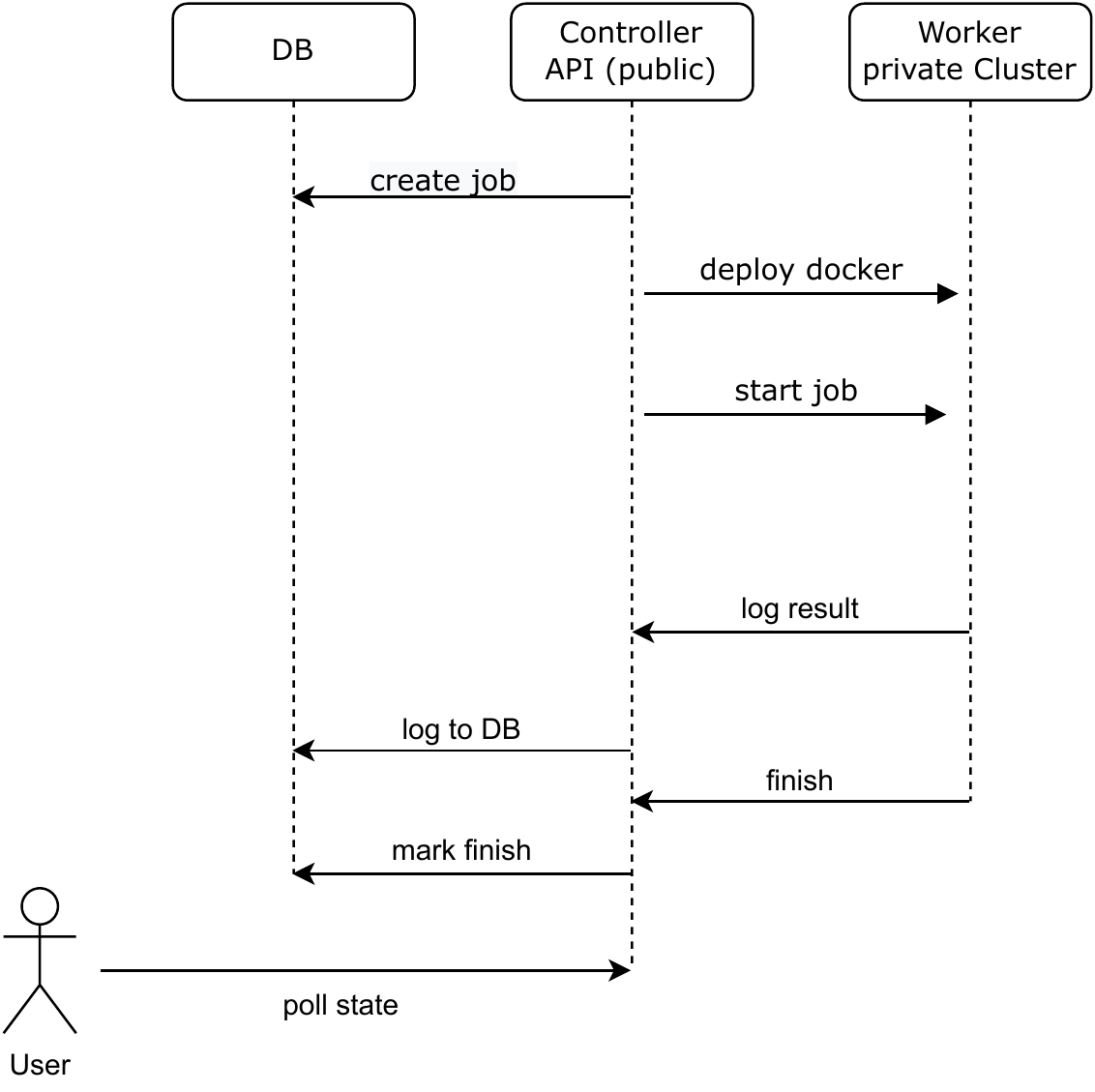}
    \caption{\textbf{(Remote) Execution in AutoDOViz:} The sequence diagram outlines the interplay between users, API controller and compute cluster workers to track RL agents, allowing the user interface to present logs, graphs and report on agent progress in real-time.}
    \label{fig:hwtl_impl}
\end{figure}

Figure \ref{fig:hwtl_impl} shows a sequence diagram of our proposed architecture towards HwtL featuring a controller plane in action. When users want to execute a job in our user interface, the controller manages workers, their execution, and logging of progress in real time.
The controller first only creates an entry about the job to be executed in the database. In return, the controller receives a job ID. The controller then deploys a Dockerized version of the AutoDO engine in a custom cluster provided by the user or executes in a shared cluster. In both cases the job ID is handed to the worker, together with an API token. The worker starts execution of the AutoDO agent search given the selected gym and configuration in the database found via the job ID. During the execution, the worker can log events via the controller in real-time. The user interface can now present logs, charts, graphs or agent policies in real-time as events stream in from the worker. This offers an substantive improvement over the classical turn-based execution of experiments where results are made available at termination of a run. The next iteration of our HwtL approach will allow for direct manipulation of running experiments.

\subsection{Gym Composer}
\label{section:composer}



\begin{figure}
    \centering
    \includegraphics[width=1\linewidth]{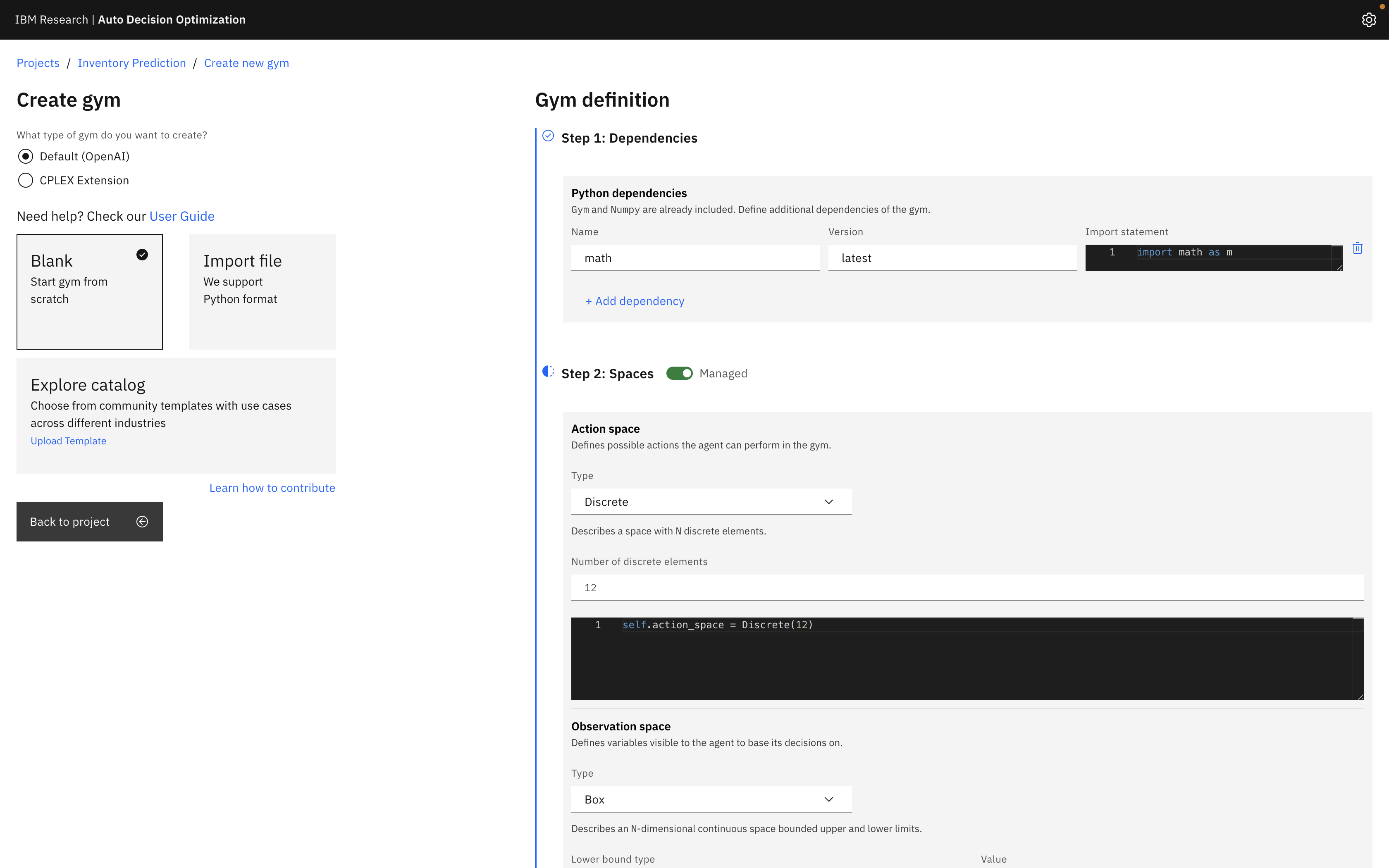}
    \caption{\textbf{Gym creation in AutoDOViz}. Users can provide OpenAI gyms in three alternative ways: (A) Compose gym via guiding Gym Composer. (B) Directly upload gym from Python file. (C) Choose gym from Gym Template Catalog.}
    \label{fig:composer}
\end{figure}

Our system greatly expands on the OpenAI \emph{gym} interface and provides a mechanism to create gyms through a composer. The composer provides several abstractions to ease creation of the gym. Specifically, definitions of observation and action space, along with reward function primitives allow the user to focus on specifying the main logic of state transitions. The system, as shown in Figure \ref{fig:composer}, aids in composing gyms using industry-specific templates for a broad range of sectors. An end-user still has several choices to make on how best to model the environment. The first choice users face is on deciding state variables, a minimal set of quantities that describe the system dynamics. Next the user is guided through the definition of a transition function, the reward function which can be comprised of a linear combination of individual metrics, as well as a termination criteria. Once the user is satisfied with their inputs to the gym composer, our API controller can auto-generate the final Python implementation script from these, which in turn serves as input to the AutoDO engine. 

\subsection{Gym Template Catalog}
\label{section:catalog}

\begin{figure}
    \centering
    \includegraphics[width=0.8\linewidth]{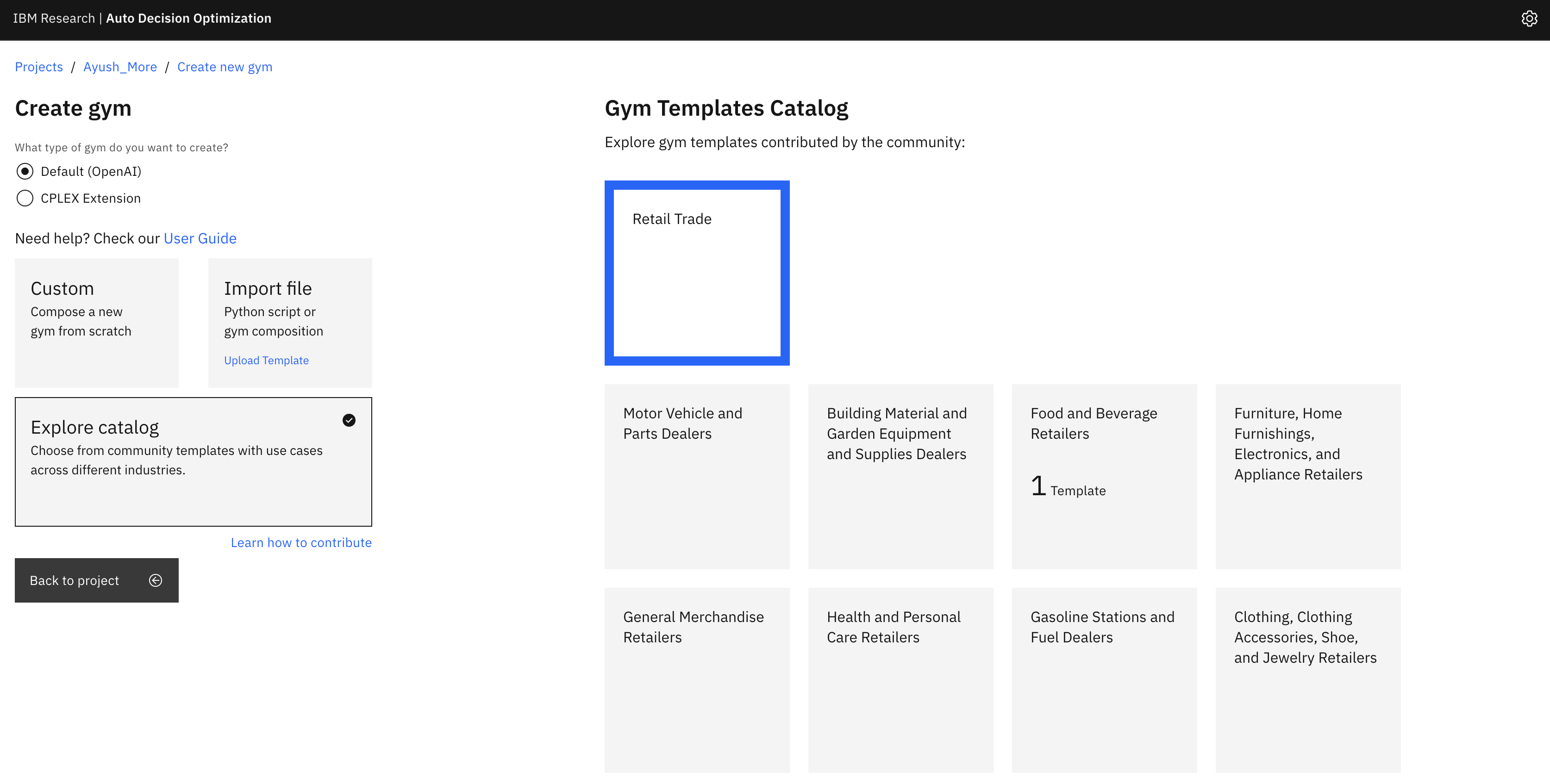}
    \caption{\textbf{Gym Template Catalog in AutoDOViz}. Gym templates can be explored optimization-type-based or industry-based (according to the NAICS standard).}
    \label{fig:catalog}
\end{figure}

With respect to \textbf{R1, R8, R9} we propose a gym template catalog. While a type-driven categorization according to \textbf{R1} is already provided by the DO researchers, we parsed options for a more domain-driven hierarchy towards \textbf{R8}. Our research on methods for classifying different kinds of business problems led us to the North American Industry Classification System (NAICS) as a robust and continually-improving breakdown of different industries. It was created jointly between the governments of The United States, Canada, and Mexico, and is updated every five years to capture emerging industries and technologies. The NAICS summarizes industries into 20 sectors, and hierarchically expands into 1,012 unique industries.\footnote{\url{https://www.census.gov/naics/}} Figure \ref{fig:catalog} shows the design of our template catalog in AutoDOViz. Users can click through the catalog hierarchy via tiles on each level, with the parent node pinned to the top with a blue border. When arriving on a leaf-level, the catalog shows the list of available templates. Upon loading a gym template from the catalog the gym composer (ref. section 4.4), the composer is prefilled with the corresponding inputs so the user can study the environment and make customized changes according to their exact optimization problem.

\subsection{State-Transition Visualizations}
\label{section:stateviz}
In order to promote trust and confidence into agents \textbf{R6} as well as to add to more generic visualization techniques \textbf{R2} we propose two types of visualizations as part of AutoDOViz. Consider a typical evaluation protocol as provided in table \ref{tab:agent_eval_proto}, consisting of chosen action, state of the environment and reward at any step in the process. Figure \ref{fig:temp_trans} then depicts a graphical representation of the evaluation protocol, with states as vertices of the graph and the trajectory of the agent shown as edges. The time step of transition is shown as labels on the edges.

\begin{minipage}{0.5\textwidth}
\centering
\begin{table}[H]
    \begin{center}
    \caption{Typical RL agent evaluation protocol.}
    \label{tab:agent_eval_proto}
    \begin{tabular}{c c c c c}
        \textbf{Step} & \textbf{Action} & \textbf{State} & \textbf{Reward} & \textbf{$\delta_\text{Reward}$}\\
        \hline
          ... & ... & ... & ... & ... \\
          $t_k$     & \textcolor{blue}{$A_1$} & \textcolor{purple}{$S_1$} & 72 &  +1 \\
          $t_{k+1}$ & \textcolor{blue}{$A_2$} & \textcolor{purple}{$S_3$} & 75 & +3 \\
          $t_{k+2}$ & \textcolor{blue}{$A_3$} & \textcolor{purple}{$S_1$} & 74 & -1 \\
          $t_{k+3}$ & \textcolor{blue}{$A_1$} & \textcolor{purple}{$S_4$} & 78 & +4 \\
          $t_{k+4}$ & \textcolor{blue}{$A_3$} & \textcolor{purple}{$S_2$} & 81 & +3 \\
          $t_{k+5}$ & \textcolor{blue}{$A_2$} & \textcolor{purple}{$S_1$} & 80 & -1 \\
          $t_{k+6}$ & \textcolor{blue}{$A_3$} & \textcolor{purple}{$S_3$} & 82 & +2 \\
          ... & ... & ... & ... & ...
    \end{tabular}
    \end{center}
\end{table}
\end{minipage}
\begin{minipage}{0.5\textwidth}
\begin{figure}[H]
\centering
\begin{tikzpicture}
    \begin{scope}[every node/.style={circle,thin,draw}]
        \node (A) at (0,0) {$S_1$};
        \node (B) at (3,0) {$S_2$};
        \node (C) at (0,3) {$S_3$};
        \node (D) at (3,3) {$S_4$};
    \end{scope}
    \begin{scope}[>={Stealth[black]}, 
        every node/.style={fill=white,circle}, 
        every edge/.style={draw=black,thin}]
        \path [->] (A) edge[bend right=-50] node {$t_{k+1}$} (C);
        \path [->] (C) edge[bend right=-45] node {$t_{k+2}$} (A);
        \path [->] (A) edge node {$t_{k+3}$} (D);
        \path [->] (D) edge node {$t_{k+4}$} (B);
        \path [->] (B) edge node {$t_{k+5}$} (A);
        \path [->] (A) edge[bend right=-5] node {$t_{k+6}$} (C);
    \end{scope}  
\end{tikzpicture}
\caption{Temporal transition graph.}
\label{fig:temp_trans}
\end{figure}
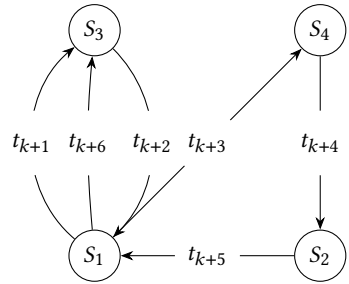
\end{minipage}

\paragraph{Transition Matrix}
Since edge labels become impractical with larger evaluation protocols, we propose a level of abstraction as shown in table \ref{tab:state_trans_mat}. Here, we infer an adjacency matrix from the evaluation protocol and count the number of the times the agent moves between any two states. A similar abstraction can be made for the action space as shown in table \ref{tab:acti_trans_mat}. While not preserving the information on sequence of steps in such transition matrices, it allows for a very compact representation of the transition pattern, which can favor comparison of many RL agents and large protocols.

\begin{minipage}{.5\textwidth}
    \centering
    \begin{table}[H]
        \begin{center}
        \caption{State transition matrix.}
        \label{tab:state_trans_mat}
        \begin{tabular}{c | c c c c }
            & \textbf{$S_1$} & \textbf{$S_2$} & \textbf{$S_3$} & \textbf{$S_4$} \\
            \hline
                \textbf{$S_1$} &  &  & \textcolor{purple}{2} & \textcolor{purple}{1} \\
                \textbf{$S_2$} & \textcolor{purple}{1} &  &  & \\
                \textbf{$S_3$} & \textcolor{purple}{1} &  &  & \\
                \textbf{$S_4$} &  & \textcolor{purple}{1} &  & 
        \end{tabular}
        \end{center}
    \end{table}
\end{minipage}
\begin{minipage}{0.5\textwidth}
    \centering
    \begin{table}[H]
        \begin{center}
        \caption{Action transition matrix.}
        \label{tab:acti_trans_mat}
        \begin{tabular}{c | c c c }
            & \textbf{$A_1$} & \textbf{$A_2$} & \textbf{$A_3$} \\
            \hline
                \textbf{$A_1$} &  & \textcolor{blue}{1} & \textcolor{blue}{1} \\
                \textbf{$A_2$} &  &  & \textcolor{blue}{2} \\
                \textbf{$A_3$} & \textcolor{blue}{1} & \textcolor{blue}{1} &
        \end{tabular}
        \end{center}
    \end{table}
\end{minipage}

\paragraph{Transition Graph}

From the abstracted adjacency matrices we can render graphs, as shown in figures \ref{fig:state_trans_graph} and \ref{fig:act_trans_graph} respectively. Here, the number of times the agent transitioned between any two states is encoded via the thickness or opacity of the edge. Note that compared to prior work, where projection of the state space happens solely based on the feature space of the environment, the proposed graph-based approach adjusts proximity of states taking into account the transitional behavior of the agent.

\begin{minipage}{0.5\textwidth}
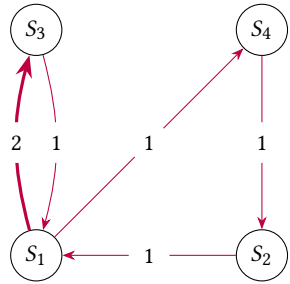
\begin{figure}[H]
\centering
\begin{tikzpicture}
    \begin{scope}[every node/.style={circle,thin,draw}]
        \node (A) at (0,0) {$S_1$};
        \node (B) at (3,0) {$S_2$};
        \node (C) at (0,3) {$S_3$};
        \node (D) at (3,3) {$S_4$};
    \end{scope}
    \begin{scope}[>={Stealth[purple]}, 
        every node/.style={fill=white,circle}, 
        every edge/.style={draw=purple,very thick}]
        \path [->] (A) edge[bend right=-15] node {$2$} (C);
    \end{scope}  
    \begin{scope}[>={Stealth[purple]}, 
        every node/.style={fill=white,circle}, 
        every edge/.style={draw=purple,thin}]
        \path [->] (A) edge node {$1$} (D);
        \path [->] (C) edge[bend right=-15] node {$1$} (A);
        \path [->] (B) edge node {$1$} (A);
        \path [->] (D) edge node {$1$} (B);
    \end{scope}  
\end{tikzpicture}
\caption{State transition graph}
\label{fig:state_trans_graph}
\end{figure}
\end{minipage}
\begin{minipage}{0.5\textwidth}
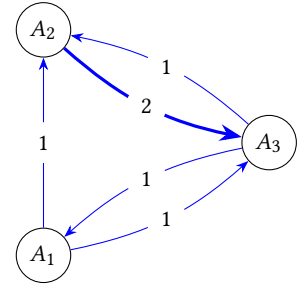
\begin{figure}[H]
\centering
\begin{tikzpicture}
    \begin{scope}[every node/.style={circle,thin,draw}]
        \node (A) at (0,0) {$A_1$};
        \node (B) at (0,3) {$A_2$};
        \node (C) at (3,1.5) {$A_3$};
    \end{scope}
    \begin{scope}[>={Stealth[blue]}, 
        every node/.style={fill=white,circle}, 
        every edge/.style={draw=blue,very thick}]
        \path [->] (B) edge[bend right=15] node {$2$} (C);
    \end{scope}  
    \begin{scope}[>={Stealth[blue]}, 
        every node/.style={fill=white,circle}, 
        every edge/.style={draw=blue,thin}]
        \path [->] (A) edge node {$1$} (B);
        \path [->] (A) edge[bend right=15] node {$1$} (C);
        \path [->] (C) edge[bend right=15] node {$1$} (A);
        \path [->] (C) edge[bend right=15] node {$1$} (B);
    \end{scope}  
\end{tikzpicture}
\caption{Action transition graph.}
\label{fig:act_trans_graph}
\end{figure}
\end{minipage}

\paragraph{Scalability} In AutoDOViz, we implemented both the matrix and graph-based technique to present behavioral information about the agent. With real-world evaluation protocols having larger action or state spaces and going into the thousands of steps per episode, our visualizations need to be able to scale well. With respect to transition matrices, where the scalability bottleneck is the dimensionality of the matrix, the user can cluster states prior to abstraction of the matrix. This scales since the matrix will then be limited to a user-defined dimensionality $k$ the number of clusters. Furthermore, states can be filtered based on certain criteria even before going into the clustering algorithm. Figures \ref{fig:matrix1} and \ref{fig:matrix2} depict a 10-means clustering of states before and after training. Initial random behavior of the agent is replaced by more selective transitions during training, as to avoid less rewarding states of the environment.

\begin{minipage}{0.5\textwidth}
\centering
\begin{figure}[H]
    \centering
    \includegraphics[width=0.45\linewidth]{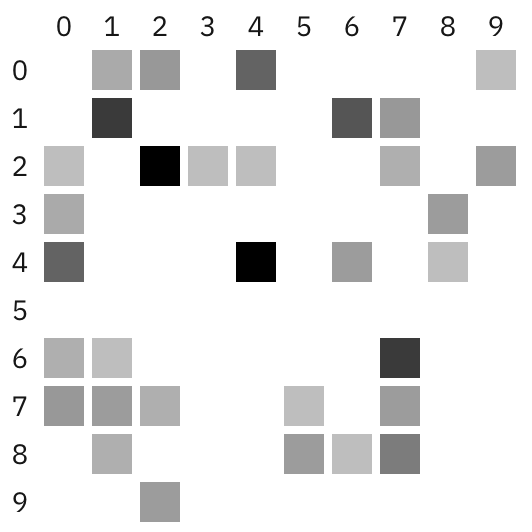}
    \caption{Clustered state transitions of early training episode.}
    \label{fig:matrix1}
\end{figure}
\end{minipage}
\begin{minipage}{0.5\textwidth}
\begin{figure}[H]
    \centering
    \includegraphics[width=0.45\linewidth]{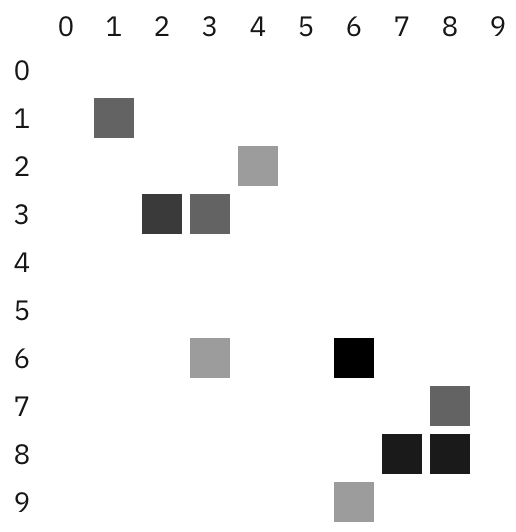}
    \caption{Clustered state transitions of late training episode.}
    \label{fig:matrix2}
\end{figure}
\end{minipage}

\begin{figure}
    \centering
    \includegraphics[width=0.7\linewidth]{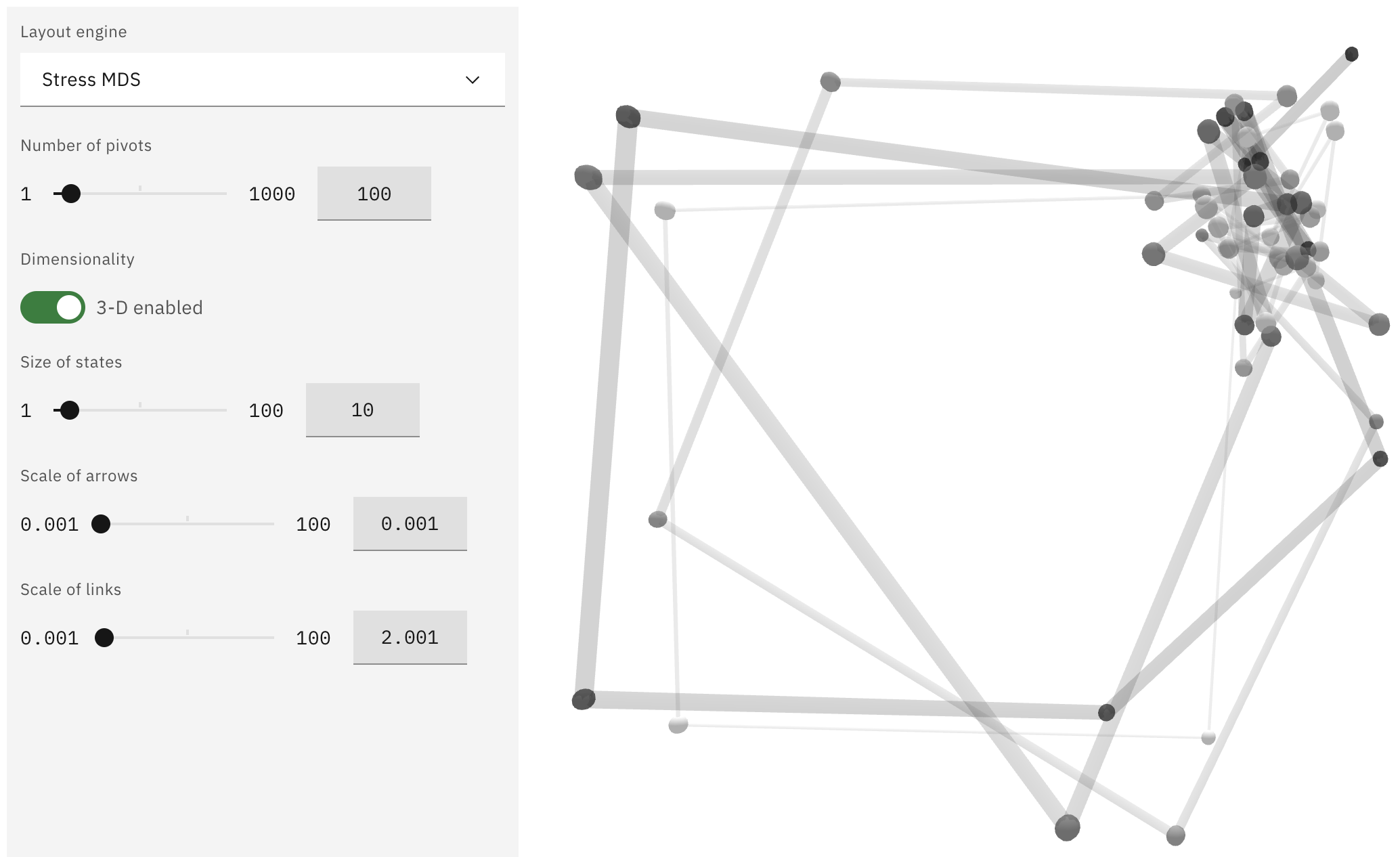}
    \caption{\textbf{Transition graph}: States are projected with multi-dimensional scaling before augmenting the tour of the RL agent.}
    \label{fig:transition_graph}
\end{figure}

\vspace{2em}
While the transition graph could in principle cope with thousands of states directly in the visualization (even without prior clustering) the bottleneck is the time to compute the projection of the states, i.e. the exact $X$,$Y$ or $X$,$Y$,$Z$ coordinates of the state nodes. For AutoDOViz, we therefore implemented state-of-the-art graph layout algorithms \cite{brandes2006eigensolver, ortmann2016sparse, zheng2018graph} in JavaScript to speedily compute layouts in the browser in 2D or 3D. Figure \ref{fig:transition_graph} shows an example where states of an evaluation protocol are projected with stress-based dimensionality reduction. We then depict the tour of the agent as it moves from state to state. Here, the thickness of an edge encodes the time of the step, so that later transitions can again be distinguished from earlier transitions.

\subsection{Policy Explanations}
\label{section:policy}
\begin{figure}
    \centering
    \includegraphics[width=1.0\linewidth]{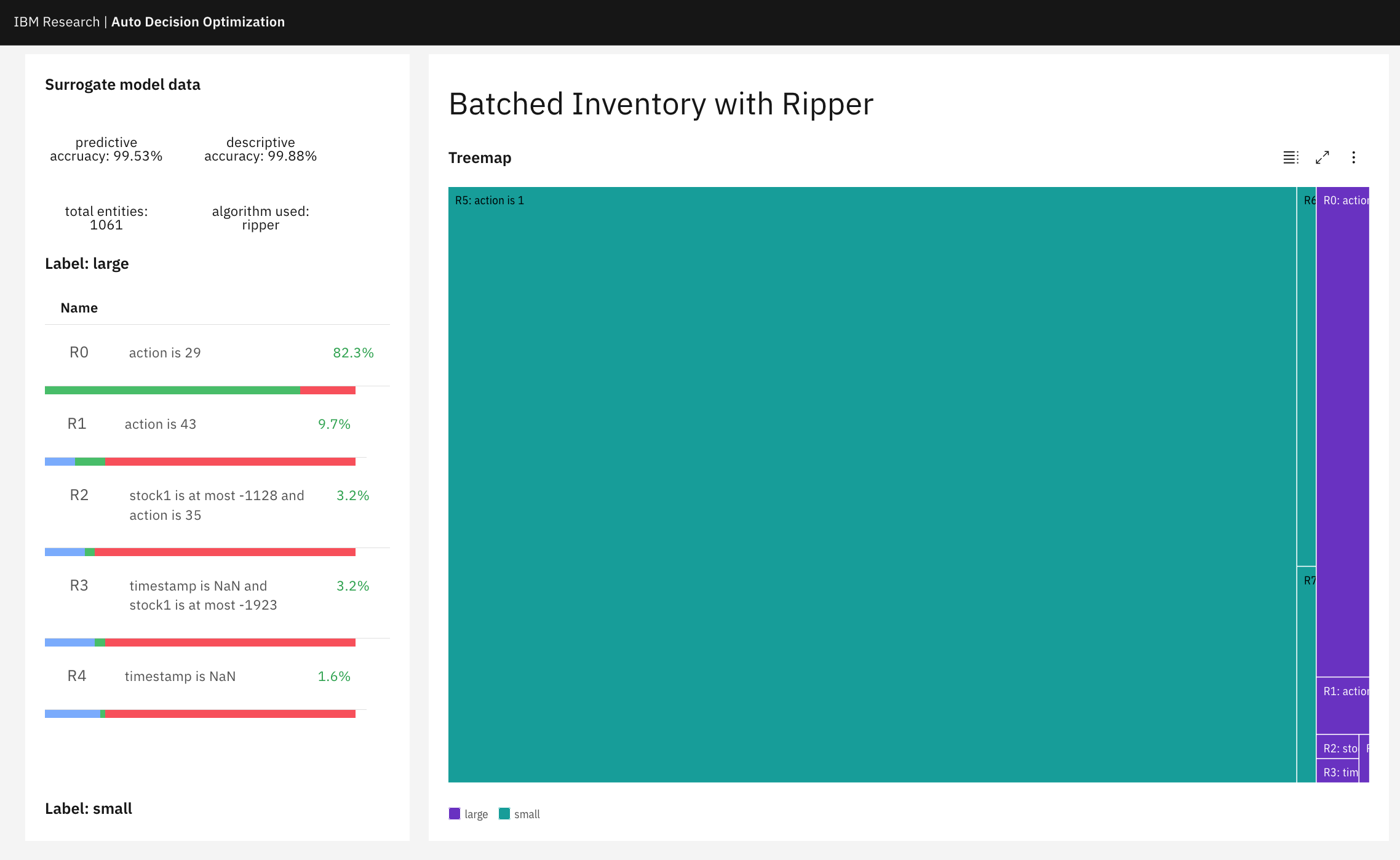}
    \caption{Extracted surrogate rule set}
    \label{fig:rule1}
\end{figure}
    
Once end-users have selected one or more agents of interest, they may inspect the agent's behavior using a surrogate rule set. The surrogate rules are learnt using observed state-action pairs of the policy behaviour in order to classify the action taken by the agent based on the state \cite{mccarthy2022boolean}. The resulting surrogate is a set of boolean logic rules that attempt to encompass and translate an agent's behavior in the most compact way possible. Rules are hierarchical, the first covers the largest population in the data, then any unexplained data points are covered by the second rule, followed by the third etc. In order to approximate an agent's behavior in the form of rules, users may choose a set of evaluations over a predefined interval of episodes, ranging from 0 to 20. The rule extraction method performs more accurately with more inputs but early episodes may not be as representative of the final agent as the final episodes. Once this interval is selected, all evaluations are concatenated into one, then users are required to select and batch a desired column into buckets as rules require a categorical prediction label. Once this batching process is completed, the selected column is discarded and rules are generated for each available prediction label. These are displayed as a list of boolean rules with their respective coverage chart as well as a global coverage tree map, displaying the weight of each rule. Boolean rules enable end-users to quickly understand an agent's behavior and compare it to other similar agents, providing better comparative insights than simple accuracy scores. 

\section{Evaluation}


\subsection{User Study with Data Scientists}

Building on results of prior expert interviews potential target end-users for AutoDOViz were identified to be data scientists. To validate this assumption and the platform's development with regard to design requirements a user study was conducted with 13 data scientists\footnote{University collaborators were not directly involved in conduction of the user study with human subjects so this study is not subject to federal IRB oversight. IBM Business Conduct Guidelines have been followed at all times \url{https://www.ibm.com/investor/governance/business-conduct-guidelines}.} employed internally in the company from across 8 different countries, with varying levels of ML experience. Participants were recruited both from pools of self-selecting volunteers via an open call 
on the company’s internal messaging software,
and others whom were contacted directly. The participants were not additionally compensated for their time. 
Participants were given no prior knowledge of study content apart from that it was to test a novel UI aimed towards data scientists. The aim of the study was to broadly evaluate whether the system successfully affords the user the ability to employ AutoDOViz to create and manage gyms and configure agents, as set out by the design requirements in section 3.3 and design choices made in section 4.  We also aimed to evaluate the UI to identify any usability concerns and to solicit feedback about potential improvements to the tool. 
The study took the form of a 45 minute online session for each participant comprising a set of tasks and a semi-structured interview, which was recorded for later analysis. Detail about the participants is given in Table~\ref{tab:prestudy_demographic}. Each participant was guided by an interviewer to complete a pre-study questionnaire, complete a list of tasks using an online environment and then complete a post-study questionnaire to reflect upon the experience. 

\paragraph{Pre-study Questionnaire}
This questionnaire contained 14 questions covering demographic information to learn more about the participants as well as questions to determine their level of prior knowledge and trust in Auto ML, reinforcement learning, and decision optimization.   

\paragraph{Tasks}
Participants were encouraged to "think aloud" as they followed through a set of tasks. In order to account for differing knowledge levels, an explanation of the concept of OpenAI gyms was given to some of the participants. When incorrect basic assumptions were made, help was provided to allow them to proceed with the tasks if necessary. Participants were asked to answer questions designed to check their ability to gather information from different parts of the UI and to assess their understanding of the scenario. After the interviews, participants' task responses were categorized as correct/successful or incorrect/fail by the interviewers, taking into account the level of additional help required to get through the tasks. 

\begin{figure}
    \centering
    \includegraphics[width=1.0\linewidth]{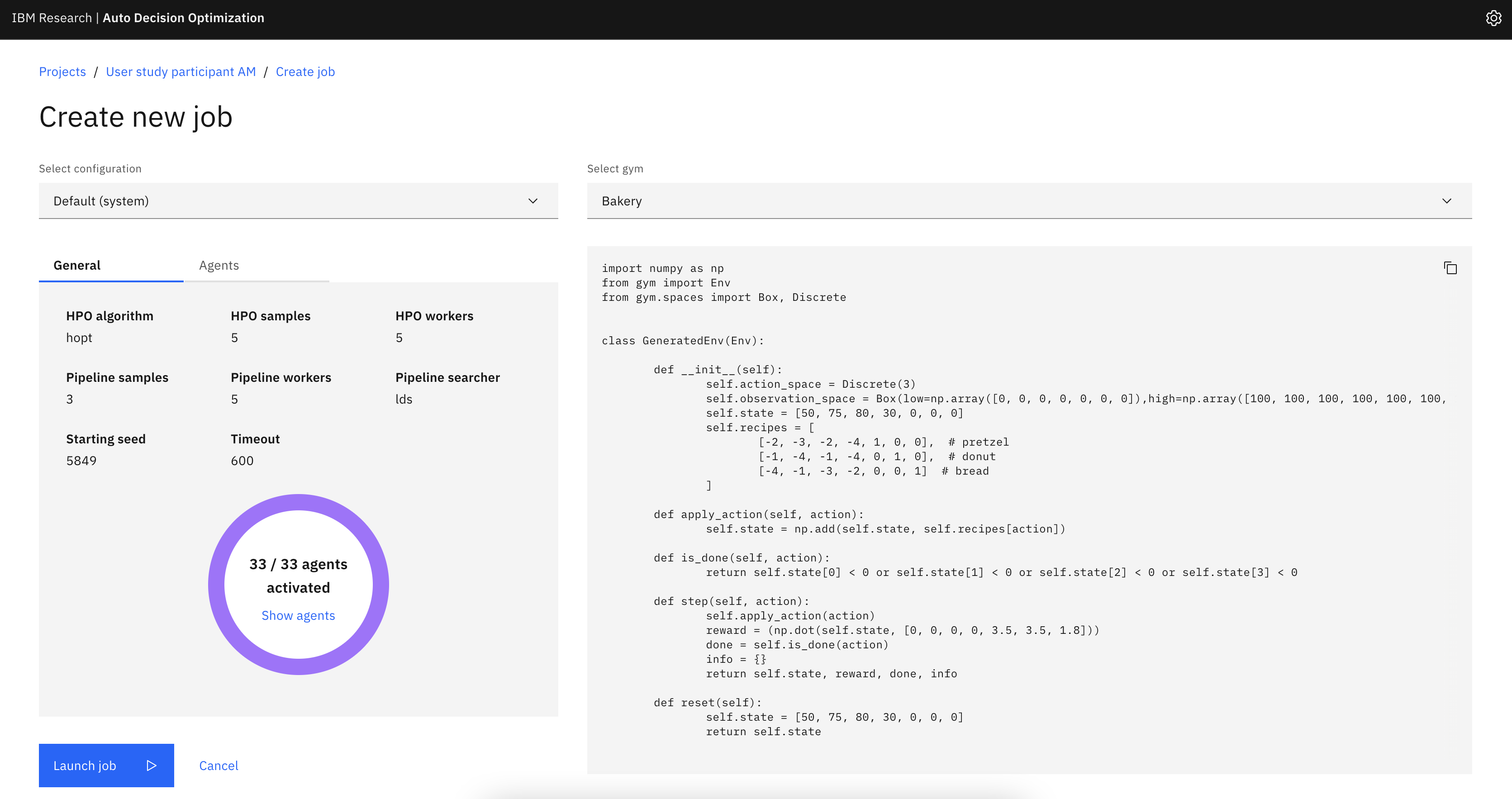}
    \caption{\textbf{Job preview screen in AutoDOViz:} Execution of a job to evaluate agents on \emph{Bakery}-gym during user study.}
    \label{fig:exec_bakery}
\end{figure}

As earlier interviews with DOs highlighted the importance of a thorough understanding of the business problem and processes rather than simply understanding through the lens of data, scenarios were prepared using real-world examples to allow assessment of both how well the UI assisted in data scientists' understanding of the business problem, as well as their success in creating, editing and configuring OpenAI gyms and jobs using the composer interface. In the first scenario, participants were asked to use the online environment and a configuration file to create and modify an OpenAI gym representing a bakery (ref. figure \ref{fig:exec_bakery}), which included details of certain recipes and how these affected inventory levels. Participants were provided with a link to an online tool and a pre-configured JSON file containing details of the bakery to use to complete the task. The questions were designed to guide them through the process of creating a gym and creating and modifying jobs and configurations, and assess their understanding of the optimization problems posed in the bakery. Participants were asked to use the catalog to load another scenario based on a produce arrangement problem of the chemical reactions between different fruits (described in section 3.2) and asked to create an additional metric which determines the distance between specified fruits.

\paragraph{Post-study reflection}
The post-study questionnaire took the form of 14 questions, including 11 5-point Likert agreement scale questions attempting to assess trust in the system, ease of use and usefulness of the UI and visualizations, attitude to the catalog and any change in their self-reported level of understanding of reinforcement learning. Free-text sections were provided for the participants to record what they liked and disliked about the tool. These were used as a basis for informal discussions with the participants to further determine their attitudes to the tool and how it might fit into their current workflows.


\subsection{Results}

In the following subsections we report results from our user study leading up to a discussion of our findings and limitations in section 6.

\subsubsection{Pre-study questionnaire}

Table \ref{tab:prestudy_demographic} shows the demographic data on user study participants. While gender is balanced (\textbf{A1}), age distribution leans towards younger professionals (\textbf{A2}). Yet, 23.1 \% of participants were above the age of 51, with 1 participant above the age of 61. All participants had a university degree, with the majority of participants having advanced degrees (\textbf{A3}), also in fields \emph{outside} of computer science, data science or artificial intelligence  (\textbf{A4}). All users had prior experience in machine learning or data science, with the majority between 2 and 10 years, and 1 outlier with up to more than 30 years of experience in machine learning (\textbf{A5}). Many (46.2 \%) users stated that they had no prior experience with decision optimization (\textbf{A6}).

\begin{table}
    \begin{center}
    \caption{Participant demographic data gathered from pre-study questionnaire}
    \label{tab:prestudy_demographic}
    \begin{tabular}{r l c c c c c}
    \hline
        \textbf{A1} & Gender      & \textbf{Female} & \textbf{Male} \\
                    & & 6 (46.15 \%) & 7 (53.85 \%) \\
        \hline
        \textbf{A2} & Age         & \textbf{20 - 30} & \textbf{31 - 40} & \textbf{41 - 50} & \textbf{51 - 60} & \textbf{61 - 70} \\
            & & 8 (61.54 \%)       & 2 (15.39 \%)       &   0 (0 \%)      & 2 (15.38 \%)       & 1 (7.69 \%)  \\
        \textbf{A3} & Degree & \textbf{Bachelor} & \textbf{Master} & \textbf{PhD} \\
        & & 4 (30.77 \%) & 6 (46.15 \%) & 3 (23.08 \%) \\
        \textbf{A4} & Field of study
                    & \multicolumn{5}{l}{Data Science \& Artificial Intelligence (3), } \\
                    & & \multicolumn{5}{l}{Computer Science (3), Physics (2),}\\
                    & & \multicolumn{5}{l}{Design Engineering (1), Electrical Engineering (1),}\\
                    & & \multicolumn{5}{l}{Geospatial Science (1), Computer Applications (1), }\\
                    & & \multicolumn{5}{l}{Tourism \& Communication (1)}\\
        \hline
        \textbf{A5} & Experience with ML/DS (years) & \textbf{0-1} & \textbf{2-4} & \textbf{5-10} & \textbf{11-30} & \textbf{30+} \\
        && 1 (7.69 \%) &  6 (46.15 \%)  &  3 (23.08 \%)  &   2 (15.38 \%)   & 1 (7.69 \%)  \\
        \textbf{A6} & Experience with DO (years) &  \textbf{0} & \textbf{1} & \textbf{2-3} \\
        & &  6 (46.15 \%) & 4 (30.77 \%) & 3 (23.08 \%) \\

        \hline
         \end{tabular}
    \end{center}
\end{table}

In self-assessment, 84.61 \% of users reported to have worked or frequently work with automated machine learning (\textbf{A7}). Furthermore, 61.5 \% of participants state to have worked with RL before (\textbf{A10}), however, only 30.8 \% claimed skills beyond beginner level (\textbf{A8}). All participants with prior RL experience have seen or directly experimented with OpenAI Gym implementations before. Less than half of the users seemed interested in applying AutoDO in real-world problems with clients (\textbf{A9}). However, 69.2 \% reported they would trust automated decision optimization (\textbf{A12}). Regarding the execution environment for RL agent optimization workers, only 1 user felt comfortable to use automated decision optimization in a shared environment, 38.5 \% preferring to exclusively work in custom environments, and 53.9 \% with the desire for both (\textbf{A13}).

\begin{table}
    \begin{center}
    \caption{Results pre-study questionnaire}
    \label{tab:prestudy}
    \begin{tabular}{r l c c c c c}
        \hline
        &  (Scale 1 to 5) & \textbf{1} & \textbf{2} & \textbf{3} & \textbf{4} & \textbf{5} \\
        \textbf{A7} & Experience AutoML   & 1(7.69 \%) & 1 (7.69 \%) & 5 (38.47 \%) & 2 (15.38 \%)  & 4 (30.77 \%) \\
        \textbf{A8} & Experience RL      & 1 (7.69 \%) & 3 (23.08 \%) & 5 (38.47 \%) & 4 (30.77 \%) & 0 (0 \%) \\
        \textbf{A9} & Interest in AutoDO & 1 (7.69 \%) & 1 (7.69 \%) & 5 (38.47 \%) & 2 (15.38 \%)  & 4 (30.77 \%) \\
        \hline
                              &  & \textbf{Yes} & \textbf{No} \\
        \textbf{A10} & Previously worked with RL & 8 & 5 \\
        \textbf{A11} & Previously used Simplex algorithm & 3 (23.08 \%) & 10 (76.92 \%) \\
        \textbf{A12} & Trust in AutoDO & 9 (69.23 \%) & 4 (30.77 \%) \\
        \hline
        \textbf{A13} & Preferred execution environment. & \textbf{Shared} & \textbf{Custom} & \textbf{Both} \\
                   & & 1 (7.69 \%)      & 5 (38.47 \%)      & 7 (53.85 \%) \\
        \hline
    \end{tabular}
    \end{center}
\end{table}

\subsubsection{Tasks}

Task \textbf{B1} was correctly executed by all but 2 users, who were unable to find the button to open their project after creation. Loading from file (\textbf{B2}) or catalog (\textbf{B3}) was mostly successful, however, with few templates in our catalog, 3 participants were confused as to why the template counter in the category tiles did not change and needed little guidance drilling down the hierarchy. Reading simple information about the gym (\textbf{B4-6}) was easy enough for most users, however 3 users gave the right answer based on reading from the wrong location: when asked about the length of the state vector they instead read the length from the observation space. The advanced task of adding a custom reward metric to a gym was successfully completed by 84.6 \% of the users. All users correctly explained the transition function, with 8 (5) people giving a very precise, problem-specific (more general) explanation. Further, 9 (5) participants provided precise, problem-specific (more general) explanations on the termination criteria, with only 1 person failing to provide a meaningful description in the given scenario. Regarding tasks to modify the configuration of the AutoDO core engine (\textbf{B10-12}), almost all users correctly solved all tasks, with only 1 person changing the number of \emph{pipeline} workers rather than the number of \emph{optimization} workers. Tasks concerned with the interpretation and launch of automation jobs (\textbf{B13-17}) were successfully executed by all participants. A single user misread the number of active agents despite looking at the right information in the tool. In \textbf{B18-19} users were presented transition matrix and transition graph visualizations and asked to answer a total of 6 questions associated with the interpretation of the visualizations: number of clusters, behavioral changes before and after training, tour of the agent. All participants gave the correct answers, with 2 different participants answering only partially correctly in some of the 6 questions. Overall, 93.1 \% of the tasks were executed correctly.

\begin{table}
    \begin{center}
    \caption{Results tasks}
    \label{tab:tasks}
    \begin{tabular}{r l r r}
        \hline
        & & \textbf{Correct/success} & \textbf{Incorrect/fail} \\
                    & \textbf{Project} \\
        \textbf{B1} & Create  & 11 & 2 \\
        \hline
         & \textbf{Gym} \\
        \textbf{B2} & Load from file & 12 & 1 \\
        \textbf{B3} & Load from catalog  & 10 & 3 \\
        \textbf{B4} & Read size of action space  & 13 & 0  \\
        \textbf{B5} & Read length of state vector  & 10 & 3 \\
        \textbf{B6} & Read reward  & 13 & 0 \\
        \textbf{B7} & Add reward metric  & 11 & 2 \\
        \textbf{B8} & Explain transition function  & 8 + 5 & 0 \\
        \textbf{B9} & Explain termination criteria  & 9 + 3 & 1 \\
        \hline
                     & \textbf{Configuration} \\
        \textbf{B10} & Change number of workers  & 12 & 1 \\
        \textbf{B11} & Enable/disable agents  & 13 & 0 \\
        \textbf{B12} & Change agent hyperparameter  & 13 & 0 \\
        \hline
                     & \textbf{Job} \\
        \textbf{B13} & Read number of workers  & 13 & 0 \\
        \textbf{B14} & Read number of agents  & 12 & 1 \\
        \textbf{B15} & List 3 agents  & 13 & 0 \\
        \textbf{B16} & Read agent hyperparameter  & 13 & 0 \\
        \textbf{B17} & Launch  & 13 & 0 \\
        \hline
                     & \textbf{Visualization} \\
        \textbf{B18} & Interpret matrix  & 12 & 1 \\
        \textbf{B19} & Interpret graph  & 12 & 1 \\
        \hline
        & \textbf{TOTAL} & \textbf{230} (93.1 \%) & \textbf{17} (6.9 \%) 
    \end{tabular}
    \end{center}
\end{table}

\subsubsection{Post-study questionnaire}
Table~\ref{tab:poststudy} shows results of the post-study questionnaire with a Likert scale from 1 (strongly disagree) to 5 (strongly agree). 92.3 \% of the participants found the gym composer to be helpful or very helpful (\textbf{C1}). 84.6 \%  strongly agreed that the visualizations provided useful insights on RL agents (\textbf{C2}). All users found it very easy to launch jobs once a gym was created (\textbf{C3}). 11 of 13 participants strongly agreed that knowing the behavior of individual metrics of reward functions made them more comfortable using an agent in production (\textbf{C4}). While not all participants felt a strong desire to \emph{contribute} to the gym template catalog (\textbf{C5}), 92.3 \% could see themselves consulting the template catalog before coding a gym themselves (\textbf{C6}). 10 participants expressed strong preference to be able to fine-tune the AutoDO engine with configurations (\textbf{C7}). 69.2 \% agreed to trust or strongly trust AutoDO, with only 4 participants being neutral. No participant reported distrust (\textbf{C8}). 9 participants expressed strong or advanced understanding of RL after our study, 2 were neutral and only 2 participants reported less than neutral understanding of RL (\textbf{C9}). 92.3 \% of users would like to use AutoDO in their own projects with only 1 user being neutral (\textbf{C10}). Lastly, 3 participants did not necessarily prefer to execute AutoDO jobs in a custom environment rather than a shared one, 5 participants were neutral, 3 would prefer to and 2 users would strongly prefer to execute AutoDO in their custom environment/private cluster (\textbf{C11}).

\begin{table}
    \begin{center}
    \caption{Results post-study questionnaire}
    \label{tab:poststudy}
    \begin{tabular}{r l r r r r r}
        \hline
         &  & \textbf{1} & \textbf{2} & \textbf{3} & \textbf{4} & \textbf{5} \\
        \hline
        \textbf{C1} & Gym Composer     & 0 & 0 & 1 & 6 &  6 \\
        \textbf{C2} & Visualizations   & 0 & 0 & 2 & 5 &  6 \\
        \textbf{C3} & Launching jobs   & 0 & 0 & 0 & 0 & 13 \\
        \textbf{C4} & Reward metrics   & 0 & 0 & 0 & 2 & 11 \\
        \textbf{C5} & Catalog contrib. & 0 & 0 & 4 & 3 &  6 \\
        \textbf{C6} & Catalog consult. & 0 & 0 & 1 & 0 & 12 \\
        \textbf{C7} & Engine config    & 0 & 0 & 1 & 2 & 10 \\
        \textbf{C8} & Trust AutoDO     & 0 & 0 & 4 & 5 &  4 \\
        \textbf{C9} & Understand RL    & 0 & 2 & 2 & 3 &  6 \\
        \textbf{C10} & Apply AutoDO     & 0 & 0 & 1 & 2 & 10 \\
        \textbf{C11} & Custom env.      & 0 & 3 & 5 & 3 &  2 \\
        \hline
    \end{tabular}
    \end{center}
\end{table}

\subsubsection{Post-survey likes, dislikes and reflection}
Participants made comments in the post-survey questionnaire relating to specific UX improvements (5 comments), and need for additional on-screen explanation for less familiar users (2 comments). Areas which they felt could be improved included user experience for small screen-size users, to reduce the amount of scrolling necessary in the composer screen while trying to remember relevant information. The agent listing screen was also something users felt could be refined. One user felt that interpretability of visualizations should be improved. Three participants reported that there was nothing in particular they disliked about the tool. Other suggestions included time sliders to replay real-time feedback on the visualizations of the agents' progress. The participants reported things they liked which can be summarized as the overall experience was reported as intuitive (5 comments), easy to work with (6 comments), customizable (4 comments) and transparent (1 comment). Participants called out the visualizations as an element that they found useful (4 comments). It was appreciated that the code and the builder were presented side-by-side and that the UI was generally fun to use. The interviewer discussed these reflections further with the participants to ensure that their feedback was being interpreted as intended. In general, the written feedback comments suggest that the tool is perceived as both useful and highly usable.

\section{Discussion}

\subsection{Trust, control \& transparency}
A surprisingly high level of trust in automated decision-making was already reported in the pre-study questionnaire (\textbf{A12}), with 69.23\% of participants in agreement that they had confidence in automated DO and 30.77\% against it. Several participants expressed reservations about the trust question as they wished to qualify that it could be circumstance-dependent. Still, it could be that as people involved daily in ML, they had some idea of the potential problems and rewards. \say{Perfectly comfortable with it. I know the pros and cons. I know what can go wrong. Uh, yeah, so that's the qualified yes.} After the study, we can see that the same amount of participants remains positive, but the remainder has moved to a neutral position \textbf{(R6)}. \say{I mean, I couldn't even see what happens in the back end, but it does seem to inspire trust}.

Participants further appreciated that the UI allowed them to customize gyms and job configuration quickly; this appeared to positively affect perception of control of the tool and its output \textbf{(R5)}. Although they have a level of trust in AI, they wish to remain able to verify and control the system's output. \say{Definitely having that control to a certain extent... that's definitely preferable.} \say{Yeah, I mean, ... especially if you're an expert user... you know how machine learning works, what kind of things you want to leave to the machine, and which type of things you want to control yourself.}

Transparency was an aspect participants felt was imperative, particularly on metrics. \say{knowing how it's measured, that makes one more comfortable.} The majority appeared to scroll through the generated gym code output to verify and understand.
\say{Just having insights into what is doing, maybe what are some of the trade-off that it is making... to get a little bit more of that,... }



\subsection{Accessibility for novice and expert users}
The initial interviews suggested that the tool should support users of multiple optimization problems \textbf{(R4)}. The tool aimed to balance wizard-style tasks for entry-level and direct code access to gyms for more experienced users. This appears to have been successful in that both low-code and code-first configurations are supported, and the participants recognized this. Some advocated for low-code solutions with even more adaptations for novice users, such as UI labels for the state. \say{Well, compared to code, it makes it much more accessible, that's for sure. If that's the alternative, I'll take this tool any time of the day for these kinds of things. So that's, I think, a big plus. It makes the code a little bit less daunting and makes it easier to use} Others were more comfortable having a flexible approach but thought that the UI would increase their efficiency and save time when creating a gym. \say{I like the flexibility of having code there.} Overall, the feedback would suggest that the UI successfully allows data scientists to learn to automate DO tasks with RL quickly. \say{Kudos to you guys! I liked it as a beginner and someone who's not used these things in my day-to-day project. I think it was it was very user-friendly.}


\subsection{Leveraging the UI as an educational tool}
We observe that participants' previous level of RL knowledge did not correlate directly to their success in completing the required tasks, or the fluency in which they could navigate the interface. The tool and the experience of completing the tasks in the user study attempt to make the complex area of RL more palatable to the novice user, as directed by requirement \textbf{R7}. This is supported by the increase in self-reported understanding of RL (\textbf{C9}) and comments from participants. \say{having had almost no experience with this, it was intuitive enough for me to follow and, you know, just get a sense of what was happening.} \textbf{(R7)}



\subsection{Collaborative potential \& resource sharing}
With respect to the gym template catalog (\textbf{R1, R8, R9}), participants seemed well-disposed towards finding and reusing code samples from the catalog for different domains. \say{As a new user, you probably want to have a look for some examples before you go in and try on yourself}. However, they were less sure that they would contribute to the catalog due to concerns about client confidentiality for their work. 

One user in particular seemed eager to collaborate in AutoDOViz with their clients and external stakeholders \textbf{(R3)}.
\say{With this tool, it's something that I would consider playing around with, coming up with use cases, you know, doing some testing. That's definitely something. Even opening this conversation with clients... and thinking of use cases in the industry could be beneficial. But definitely, this is this could be a conversation starter for sure}.

Several participants asked when AutoDOViz would be integrated into their existing toolkits for them to start to experiment with it for clients.
\say{I'm a data scientist within client engineering, so we work on... 6 to 8-week intervals with their clients and create a product within that, an MVP or proof of concept. So definitely, something like this, I can see, is very helpful, rather than me focusing on, like, the little itty-bitty like the parameters. I think you should have a Kaggle mindset to see how you can make the best model, but to have something that works and the client's happy with it. I think this is a good option.}

When questioned about attitudes to working in a shared versus a custom environment (\textbf{C11}), it emerged that this is highly use-case dependent \textbf{(R1, R8)}, on customer need driven by security, privacy, and cost factors. Participants working in pre-sales on customer demos would have a different perspective than those working on government projects.

\subsection{Design decisions \& mental model}
The UI was designed and developed using a UI framework that was familiar to the data scientists from other software that they used daily \textbf{(R2)}. This consistency was appreciated by some of the participants and they suggested that it helped them navigate through the tasks more smoothly.
\say{Right and the placement also, right? I mean, yeah, it helps to... it's sort of like muscle memory that okay I know where to find.} 

\say{As a whole, it kind of fits my mental model. I think of where you would click and where you would find things.  So, that fits well as well...it's kind of logical that way.}

\subsection{Limitations}



Although the population was diverse in age and experience level, all participants were selected from the same company, which might yield familiarity with similar tools. This could potentially affect the generalization of results if the study was replicated with external participants (external validity).
Overall, we could criticize differing ML/RL knowledge levels amongst participants, such as using a task script augmented by verbal instructions and discussion, which differed slightly for each participant. Also, the participants are subjectively self-reporting their skill level (internal validity).
The tasks and examples chosen could be simpler than those encountered in real-world scenarios. However, we needed to balance ease of access for a beginner and an experienced user (construct validity). Another limitation could be the question of how to measure trust in AutoDOViz. Yes-No questions may not provide the nuance that participants expressed verbally since there were a few "Yes, but..." responses. 


\section{Conclusion and Future Work}

We presented AutoDOViz, an interactive user experience for automated DO with RL. Findings from semi-structured expert interviews with DO practitioners and business consultants generated design requirements for human-centered automation for DO. We implemented a system design and evaluated our software in a rigorous user study with data scientists. We found that they are significantly more open to engaging in DO after using our proposed solution. AutoDOViz promotes trust and confidence in RL agent models via reward metrics and makes the automated training and evaluation process more accessible and understandable. This is further expanded by leveraging the benefits of AutoDO algorithms for RL pipeline search to generate policy insights and advanced visualizations to enable communication between DO experts and domain experts. The novel gym composer and proposed gym template catalog lower the barrier of entry for data scientists in problem specification for RL problems. The provided streaming architecture enables Human-within-the-Loop for enhanced interaction, with real-time \emph{manipulation} subject to future work. The user study further gave rise to helpful design improvements in the overall user interface. Future work will be conducted in the direction of real-world applications for our proposed system to test and demonstrate the interface's efficiency amongst various personas. In this regard, we would also like to iterate on the application to become role-based and add collaborative features directly to the interface. Lastly, more problem-specific visualizations will make the agents' performance more interpretable and explainable.


\bibliographystyle{ACM-Reference-Format}
\bibliography{sample-base}

\appendix

\end{document}